\documentclass[aps,pre,reprint,showpacs,onecolumn,superscriptaddress]{revtex4-1}
\usepackage{graphicx}
\usepackage{dcolumn}
\usepackage{bm}
\usepackage{amsmath}
\usepackage{graphicx}
\usepackage{amsfonts}
\usepackage{amssymb}
\usepackage{color}
\usepackage{tikz}
\usepackage{changes}
\definechangesauthor[name={Per cusse}, color=orange]{per}
\setremarkmarkup{(#2)}
\usepackage{enumerate}
\usepackage[normalem]{ulem}

\def\mean#1{\left \langle#1\right \rangle}
\def\tuple#1{\{#1\}}
\newcommand{\av}[1]{\left\langle #1 \right\rangle}
\newcommand{\bracket}[1]{\left(#1\right)}
\newcommand{\N}{\mathcal{N}}
\newcommand{\M}{\mathcal{M}}
\newcommand{\Nh}{\mathcal{H}}
\newcommand{\J}{\mathcal{J}}
\newcommand{\lp}{\mathit{l}}
\newcommand{\D}{\mathcal{D}}
\newcommand{\Ss}{\mathcal{S}}
\newcommand{\R}{\mathcal{R}}
\newcommand{\p}{\mathcal{P}}
\newcommand{\Cv}{\mathcal{C}}

\usepackage[toc,page]{appendix}
\graphicspath{{Figures/}}
\usepackage{comment}
\newcommand{\Dunits}{($\mu$A/cm$^2$)$^2$\text{ms}}
\newcommand{\IAH}{J_{\text{AH}}}

\newcommand{\Ith}{J_{\text{th}}}
\newcommand{\Ieff}{J_{\text{eff}}}
\newcommand{\Deff}{S_{\text{eff}}}
\newcommand{\Ieffk}{J_{\text{eff},k}}
\newcommand{\Deffk}{S_{\text{eff},k}}

\graphicspath{{Figures/}}
\newcommand{\red}[1]{{\color{red}{#1}}}

\def\mean#1{\langle#1\rangle}

\begin{document}

\title{Variability of collective dynamics in random tree networks of strongly-coupled stochastic excitable elements}
\author{Ali Khaledi-Nasab}
\email{ali.khaledi1989@gmail.com}
\affiliation{Department of Physics and Astronomy, Ohio University, Athens, Ohio 45701, USA} 
\affiliation{Neuroscience Program, Ohio University, Athens, Ohio 45701, USA}
\author{Justus A. Kromer}
\email{jkromer@stanford.edu}
\affiliation{Stanford University, Department of Neurosurgery, Stanford, CA, 94305, USA}
\author{Lutz Schimansky-Geier}
\email{alsg@physik.hu-berlin.de}
\affiliation{Department of Physics and Astronomy, Ohio University, Athens, Ohio 45701, USA} 
\affiliation{Department of Physics, Humboldt-Universit\"at zu Berlin, Newtonstrasse 15, 12489 Berlin, Germany}
\affiliation{Bernstein Center for Computational Neuroscience, Berlin, Germany}
\author{Alexander B. Neiman}
\email{neimana@ohio.edu}
\affiliation{Department of Physics and Astronomy, Ohio University, Athens, Ohio 45701, USA}
\affiliation{Neuroscience Program, Ohio University, Athens, Ohio 45701, USA}

\date{\today}


\begin{abstract}
We study the collective dynamics of strongly  diffusively coupled excitable elements  on small random tree networks.
Stochastic external inputs are applied to the leaves causing large spiking events. Those events  propagate along the tree branches and, eventually, exciting the root node. Using Hodgkin-Huxley type nodal elements, such a setup serves as a  model for sensory neurons with branched myelinated distal terminals. We focus on the influence of the variability of tree structures on the spike train statistics of the root node. We present a statistical description of random tree network and show how the structural variability translates into the collective network dynamics.
In particular, we show that in the physiologically relevant case of strong coupling the variability of collective response is determined by the joint probability distribution of the total number of leaves and nodes. We further present analytical results for the strong coupling limit in which the entire tree network can be represented by an effective single element.
\end{abstract}


\maketitle

\section{Introduction}
The study of the dynamical properties of complex networks of nonlinear elements  \cite{newman2010networks,barabasi2016network} is an important  trend in nonlinear science \cite{boccaletti2006complex,arenas2008synchronization}. In particular, networks of coupled stochastic excitable elements are commonly used as model systems for a wide range of natural phenomena, such as pattern formation in chemical reactions \cite{kiss2003chemical,mikhailov2012engineering}, the dynamics of gene regulatory networks \cite{farkas2002social,borgatti2009network,newman2002spread,chen2015emergent}, the electrical activity of single \cite{dc2005,kinouchi2006optimal,gollo2013single} neurons and large neuronal populations \cite{bullmore2009complex}.

Network topology strongly influences the collective dynamics of coupled excitable elements \cite{boccaletti2006complex,ocker2017statistics}. For instance, coherent collective oscillations can emerge for certain coupling strengths or particular choices of network connectivity  \cite{lin2000,sagues2007spatiotemporal}. Furthermore, the emergent correlated activity of large recurrent neural networks can be linked to their connectivity \cite{ocker2017linking}. The sensitivity of complex networks to external signals and the dynamic range of the network's collective response can be maximized for the so-called critical networks \cite{chialvo2010emergent}, i.e being on the verge of a phase transition. This criticality can be achieved either by tuning the coupling strength and signal propagation parameters in the network \cite{kinouchi2006optimal,gollo2013single} or by tuning its topology \cite{larremore2011predicting}. 

The majority of works in this area are devoted to large networks with well-defined statistical properties such as degree distributions and spectra of the adjacency or Laplace matrices \cite{newman2002spread,boccaletti2006complex,arenas2008synchronization}. Due to the large number of nodes and interconnections one can average over the network structure. Then, the emergent collective dynamics can be related the statistical properties of the network's architecture \cite{buice2010systematic,ocker2017linking,ocker2017statistics}.
The situation is different when the collective dynamics of small random networks is studied. Although the network topology can be specified in terms of statistical properties, such as degree distributions, etc. , individual network realizations may differ significantly. In consequence, a detailed analysis of the relation between the collective dynamics and statistical properties of the network topology require studies of ensembles of networks realizations.

\begin{figure}[h]
	\centering 
	\includegraphics[width=0.7\columnwidth]{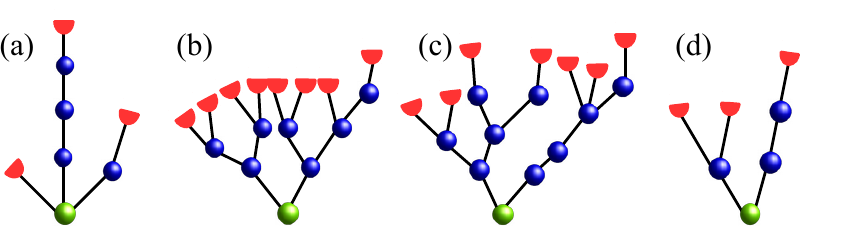}
	\caption{Examples of connectivity of nodes of Ranvier in myelinated branching terminals  of muscle spindle afferents (a--c) from \cite{banks1997pacemaker,banks1982form} and of a touch receptor afferent (d) from \cite{lesniak2014computation}. Red semicircles represent heminodes (leaf nodes)  which receive external inputs. Internal nodes of Ranvier are marked by blue circles; the  green circle marks the root node.} 
	\label{tree_sample1}
\end{figure}

In the present paper we focus on a class of small networks: small random trees of strongly-coupled stochastic excitable elements. Such networks serve as models for certain types of peripheral sensory neurons whose morphology includes tree-like branched myelinated distal terminals. Examples of such sensory neurons include cutaneous mechanoreceptors \cite{lesniak2014computation,marshall2016touch}, pain receptors \cite{besson1999neurobiology}, mechanoreceptors  in lungs (pulmonary afferents)  \cite{yu2004single,lee2014sensory}, some electroreceptors \cite{rogers2013sensory}, and muscle spindles \cite{quick1980anatomical,bewick2015mechanotransduction}. Terminal branches of such neurons are wrapped by myeline which is interrupted at the nodes of Ranvier, located at branching points.  Figure \ref{tree_sample1} exemplifies such networks for terminal branching of muscle spindle afferent neurons and for a touch receptor afferent. Starting from a primary node (green circle), branching continues for a few generations (2 -- 5), terminating at leaves, called heminodes, at which myeline ends. Heminodes receive sensory inputs from thinner neurite processes. In response, the sensory neuron generates a sequence of action potentials. 
The number of nodes and heminodes, their connectivity and the size of terminals vary among individual neurons.

Due to the high density of voltage-gated Na$^{+}$ ion channels at the heminodes, an action potential (AP) can be triggered at any heminode. Therefore, such neurons obey multiple stimulus encoding zones \cite{quick1980anatomical}. Furthermore,  despite the inevitable randomness of input signals to the individual spatially-separated terminal endings, these sensory neurons often exhibit pacemaker-like activity, characterized by noisy periodic spiking \cite{banks1997pacemaker}. Several dynamic mechanisms were proposed in order to explain AP generation, the periodicity of firing, and the observed nonlinear responses of these neurons. Those mechanisms include random mixing \cite{eagles1974afferent}, nonlinear competition between multiple pacemakers, associated with heminodes \cite{banks1997pacemaker}, and additional mechanical coupling between sensory receptors \cite{carr1998summation}. In an alternative approach, the low resistance of myelinated segments, interconnecting the individual nodes of Ranvier, leads to strong coupling of their activity.  In consequence, the stochastic firing of heminodes and nodes is synchronized and the whole branched terminal can be viewed as a single effective excitable system, which produces the corresponding firing statistics \cite{kroller1990study}. This proposal was supported by modeling studies using star \cite{justus2016} and regular tree \cite{kromer2017emergent} networks of stochastic excitable elements. In particular, in \cite{kromer2017emergent}, a strong-coupling theory was developed, allowing prediction of the firing rate and spike train variability of strongly coupled excitable elements. 

Reconstructions of myelinated terminals of sensory neurons revealed that their tree  structures varies among neurons \cite{banks1997pacemaker,banks1982form,lesniak2014computation}, see Fig.~\ref{tree_sample1}. 
This gives rise to a description of those terminals using random tree networks as models. Within this paradigm the specific coupling structure in a single myelinated terminal is just one possible realization of a random branching process, which generates random tree networks with certain statistical properties. Those properties can, for instance, be specified by providing a branching probability mass function, which, back in the experimental setup, would characterize myelinated terminal of a certain kind of neuron. As terminals of individual neurons may differ significantly, this raises the question of how this structural variability affects the statistics of neuronal firing \cite{lesniak2014computation}. 

The present paper is organized as follows. In Sec.\ref{voltage_sec} we describe a  model of Hodgkin-Huxley type excitable elements coupled on random tree network.  
The deterministic dynamics and measures of spike train variability for a tree network are described in Sec. II~B,C. In Section II~D we introduce a statistical description of random tree networks. The latter is then applied for particular examples of random binary trees in Sec.III~A. Section III~B and Appendix\ref{app:Averaging_Dynamics_Over_Shells} are devoted to the strong coupling theory, which is  applied to three examples of random trees in Sec.III~C.
We end with our conclusions in Sec. IV.

\section{Model and methods}

In the present paper, we study the collective dynamics of excitable elements located at the branching points of random tree networks. Elements are interconnected by passive branches. This setup is illustrated in  Fig.~\ref{tree_sample1}. 
\subsection{Hodgkin-Huxley type model}
\label{voltage_sec}
We assume that all nodes and passive links are identical, except for the leaf nodes, representing heminodes, which receive external inputs. Given a tree with $\N$ nodes, the dynamics of the nodes' membrane potentials are approximated by a discrete cable model \cite{ermentrout2010foundations} in which the membrane potential of the $n$th node is governed by
\begin{equation}
\label{HHmodel.eq1}
C \dot{V}_n = - I_\text{ion} + \kappa \sum_{j=1}^\N A_{n,j}(V_j - V_n) + \J_n(t).
\end{equation}
Here the index $n=1,2,...,\N$ marks the respective node. In particular, $n=1$ refers to the root node. In Eq. (\ref{HHmodel.eq1}) $C$ is the nodal membrane capacitance and the term $I_\text{ion}$ represents the nodal ionic currents.
In the following, we use a Hodgkin-Huxley type (HH) model for the ionic currents of nodes of Ranvier, which is a simplified version of the model used in \cite{mc2002}. Ionic currents are represented by 
Na$^+$ and leak currents, $I_\text{ion}=I_\text{Na}+I_\text{L}$, \cite{justus2016,kromer2017emergent}. The Na$^+$ current is 
$I_\text{Na} = g_\text{Na} m^3 h (V-V_\text{Na})$, 
where $g_\text{Na} = 1100$~mS/cm$^2$ is the maximal value of the sodium conductance and $V_\text{Na}= 50$~mV is the Na$^+$ reversal potential.
The gating activation and inactivation variables obey the dynamics 
\begin{equation}
\label{HHmodel.eq2}
\dot{m} = \alpha_m(V) (1-m)-\beta_m(V)m, \quad 
\dot{h}  = \alpha_h (V)(1-h)-\beta_h(V)h, 
\end{equation}
with the following rate functions:
\begin{eqnarray}
&&\alpha_m(V) = 1.314(V+20.4)/[1-\exp[-(V+20.4)/10.3]],\cr
&&\beta_m(V)  = -0.0608(V+25.7)/[1-\exp[(V+25.7)/11]], \cr
&&\alpha_h(V) = -0.068(V+114)/[1-\exp[(V+114)/11]], \cr
&&\beta_h(V)  = 2.52/[1+\exp[-(V+31.8)/13.4]]. \nonumber      
\end{eqnarray}
The leak current is given by  $I_\text{L}=g_\text{L}(V-V_\text{L})$ with 
$g_\text{L}=20$~mS/cm$^2$, $V_\text{L}=-80$~mV, and the nodal capacitance is set to
$C=2$~$\mu$F/cm$^2$. 

In Eq. (\ref{HHmodel.eq1}) the coupling between nodes is described by $\kappa \sum_{j=1}^N A_{n,j}(V_j - V_n)$, where $\mathbf{A}$ is the adjacency matrix of the undirected rooted tree  graph. It is a $\N\times \N$ symmetric matrix with elements $A_{i,j}=1$ for connected nodes $i$ and $j$, and $A_{i,j}=0$ for unconnected nodes, see Appendix \ref{rand_adj_matrix.apdnx} for more details. In the following the coupling strength, $\kappa$, is used as a control parameter. However, the physiologically-relevant values of $\kappa$  can be estimated from the sizes of a node, the myelinated links, and the axoplasmic resistivity \cite{ermentrout2010foundations}, giving 
the range of  $\approx$ 125 -- 1500 mS/cm$^2$ \cite{justus2016,kromer2017emergent}.
The external currents $\J_n$ are applied to the leaves only and consist of a constant and a noisy part, i.e 
\begin{equation}
\label{HHmodel.eq3}
\J_n(t) = \delta_{n,l} [J + \sqrt{2S} \, \xi_l (t)],
\end{equation}
where $l$ denotes indices of  leaf nodes; $\delta_{n,l}$ is the Kronecker delta. 
The zero-mean Gaussian white noise $\xi_l(t)$ with intensity $S$ is uncorrelated for different leaves, i.e. $\mean{\xi_i(t)}=0$, 
$\mean{\xi_{i}(t)\xi_{j}(t+\tau)}=\delta_{i,j} \, \delta(\tau)$.
Thus, leaf nodes receive random uncorrelated inputs. Other sources of noise, e.g. due to fluctuations of nodal ion channel conductances are neglected. 
In contrast to regular tree networks where inputs are administered only to nodes in the last generation of a tree \cite{kromer2017emergent}, leaves can occur at any generation in random tree networks, see Fig.~\ref{tree_sample1}. 

Eqs.(\ref{HHmodel.eq1}--\ref{HHmodel.eq3}) were integrated numerically using the explicit Euler - Maruyama method with time step of $0.1$ $\mu$s for 60 -- 600~ seconds long simulation runs. 

\subsection{Deterministic dynamics}
\label{measures.det}

We first discuss repetitive action potential generation at the root node for deterministic input currents, $S=0$ in Eq.~(\ref{HHmodel.eq3}). 
Qualitatively different dynamical operation modes of the root node are separated by a threshold value, $\Ith$, of the constant current,  $J$, applied to 
leaf nodes. While APs evoked at the leaf nodes do not fire up the root node for low currents, values of $J$ exceeding $\Ith$ result into sustained periodic firing of the root node. This is reminiscent of the dynamic behavior of a single isolated node. 
The latter is at resting state in the absence of external input, $\J=0$. A sufficiently high constant current results in an Andronov-Hopf bifurcation of the equilibrium state rendering an isolated node to fire a periodic sequence of action potentials (APs). For a single node this Andronov-Hopf bifurcation  occurs at $\IAH\approx 29.06~\mu$A/cm$^2$ \cite{kromer2017emergent}. 

\begin{figure}
	\includegraphics[width=0.7\columnwidth]{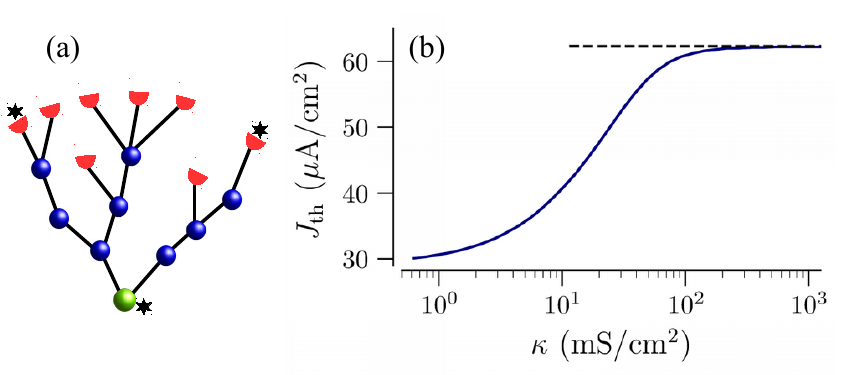}
	\caption{Sufficiently-strong input current to leaf nodes fires up the root node. (a): A sample tree with $\N=17$ nodes. The $\Nh=8$ leaf nodes are marked by red semicircles. The root node is marked by the green circle. Star symbols point at "recording" sites of voltage traces shown in Fig.~\ref{firing_ex.fig}b. The shown tree structure reproduces a reconstruction of an experimentally-observed muscle spindle afferent neuron presented in Ref. \cite{banks1982form}.
 (b): Threshold current, $\Ith$, for the onset of repetitive firing of the root node, as a function of the coupling strength, $\kappa$, for the tree shown in panel (a).  The dashed horizontal line marks the  theoretical estimate of the threshold current in the strong coupling limit, $J_\infty = (\N/\Nh) J_\text{AH} = 61.75~\mu$A/cm$^2$, see in Sec.~\ref{strong_cpl.sec}.}
	\label{sample_tree.fig}
\end{figure}

As in Ref. \cite{kromer2017emergent}, we numerically calculated the threshold current $\Ith$, which is the minimum constant current applied to the leaf nodes, which for a given coupling strength, results in the repetitive sustained generation of full-size APs (a voltage spike of at least 60~mV magnitude) at the root node. As for regular trees \cite{kromer2017emergent}, the threshold current depends on the coupling strength, $\Ith(\kappa)$, as exemplified in Fig.~\ref{sample_tree.fig}b.
As the coupling strength increases, more input current is required to sustain firing of the root node. In consequence, the threshold current increases with $\kappa$ and, finally, saturates at the limiting value $J_\infty:=\lim_{\kappa \to \infty} \Ith(\kappa)$ for strong coupling. The strong coupling regime spans the range of physiologically realistic values, $\kappa > 100$~mS/cm$^2$, for branched myelinated terminals of sensory neurons  \cite{justus2016,kromer2017emergent}.

For a given value of $\kappa$ the root node shows a sustained sequence of APs, if values of the input current are above $\Ith(\kappa)$, shown in Fig.~\ref{sample_tree.fig}b, and no APs if the value of $\J_n$ is below that curve. These two regimes are referred to as oscillatory and excitable, respectively, in the following. Note that for weak coupling the dynamics of the tree can be quite complex, e.g. not every AP generated at leaf nodes may propagate all the way to the root node. Such regimes will be studied elsewhere. 

For strong enough coupling and sufficient input current, all nodes in the tree are synchronized and fire periodic sequences of APs.  
As in the case of regular trees, the limiting value of the threshold current for strong coupling, $J_\infty$, matches the threshold value of the current of an effective single node with parameters re-scaled by the ratio of the number of inputs (leaf nodes) and the total number of nodes as, $J_\infty = (\N /\Nh) \IAH$, see dashed line in Fig.~\ref{sample_tree.fig}b. This value is derived in Sec.~\ref{strong_cpl.sec} and Appendix~\ref{app:Averaging_Dynamics_Over_Shells}.

\subsection{Spike train statistics}
\label{measures.spk}
In the presence of noise, the AP generation becomes stochastic. We are particularly interested in statistical properties of spike trains generated at the root node. We extracted a sequence of spike times of the root node, $\{t_i\}$, from $60-600$~s long simulation runs. The sequence of interspike  intervals (ISIs), $\tau_i =t_{i+1}-t_i$, is characterized by the firing rate, $r$, and by the coefficient of variation (CV), $\Cv_\tau$, 
\begin{equation}
r = \left(\overline{\tau}\right)^{-1}, 
\quad \sigma^2_\tau = \overline{\tau^2} - (\overline{\tau})^2, \quad \Cv_{\tau}=  r \sigma_\tau, 
\label{mfr.eq}
\end{equation}
where $\overline{\tau}$ and $\sigma_\tau$ are the mean and the SD of the sequence of ISIs, respectively. The bar stands for the averaging  over all ISIs in the spike train of the root node. The CV  quantifies the ISI variability. 

\begin{figure}[h!]
	\includegraphics[width=0.7\columnwidth]{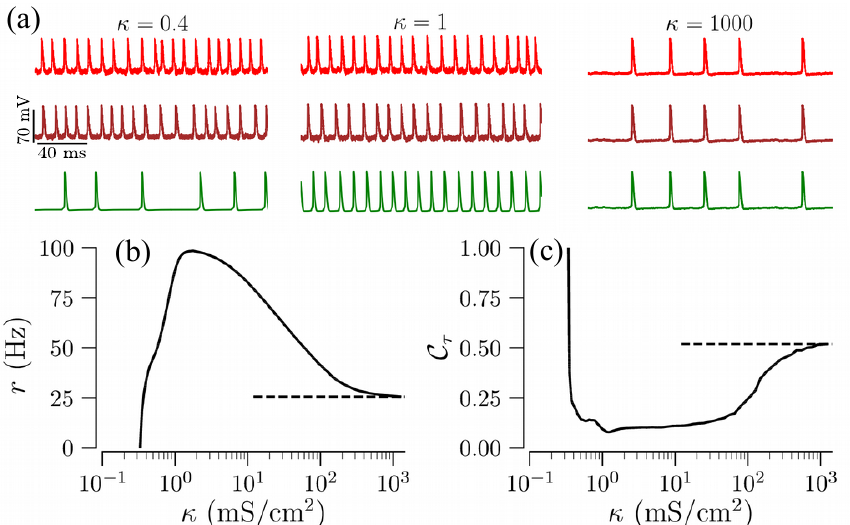}
	\caption{Stochastic dynamics of excitable elements coupled on the tree network shown in Fig.~\ref{sample_tree.fig}a.
		(a): Time traces of the membrane potentials of two leaf nodes and the root marked by stars in Fig.~\ref{sample_tree.fig}a, for the indicated coupling strengths.
		(b,c): Firing rate, $r$, and coefficient of variation, $\Cv_{\tau}$ for spike trains generated at the root node as a function of coupling strength. Dashed lines in panels (b) and (c) refer to theoretical estimates of the firing rate and CV in the strong coupling limit.
		The parameters of input currents to leaves are: $J=50$~$\mu$A/cm$^2$, $S=500$~\Dunits. 
	}
	\label{firing_ex.fig}
\end{figure}
Figure~\ref{firing_ex.fig} exemplifies the stochastic firing dynamics for the tree shown in Fig.~\ref{sample_tree.fig}a. 
The constant current was chosen such that the tree is in the excitable regime for $\kappa>20$~mS/cm$^2$.
As in regular trees \cite{kromer2017emergent}, the firing rate, $r(\kappa)$, depends non-monotonously on the coupling strength. For weak coupling, spikes generated at the leaves often fail to propagate to the root, leading to its sparse firing, Fig.~\ref{firing_ex.fig}a. This results in low values of the firing rate and large values of the CV. Furthermore, the root and leaf nodes fire asynchronously.  Increasing the coupling strength leads to stronger interaction between nodes. In consequence, synchronous coherent firing with large firing rates and small CVs emerges, see middle panel of Fig.~\ref{firing_ex.fig}a. 
As the coupling increases, more input current is required to sustain the network firing, see Fig.~\ref{sample_tree.fig}b.  Consequently, for constant input current the firing rate decrease again.
Competition of these two tendencies results in a maximum of the root node firing rate as a function of the coupling strength.
For strong coupling, nodal firing is perfectly synchronized, see rightmost traces in Fig.~\ref{firing_ex.fig}a, and the firing rate and the CV saturate at their limiting values, indicated by dashed lines in Fig.~\ref{firing_ex.fig}b,c. In Sec.~\ref{strong_cpl.sec} we show that in the case of strong coupling, a tree can be well represented by its root node which receives effective input with rescaled constant current and noise intensity.

\subsection{Statistical description of random trees networks}
\label{stat_discr.sec}
An ensemble of random trees can be constructed as a set of realizations of a stochastic branching process \cite{drmota2009random}. 
In this paradigm, the tree shown in Fig.~\ref{sample_tree.fig}a is just one possible realization of a random tree network. Each network realization causes certain statistical properties of its root node's  firing characteristics, such as those depicted in Fig.~\ref{firing_ex.fig}. 

In each realization, branching starts at the root and continues up to a prescribed maximal number of generations, $G$, or until all branches end in leaf nodes. Each node in a certain generation $g$ is located at the same path distance, $g$, from the root. The latter defines the generation $g=0$. Furthermore, each node in the $g$-th generation, $g<G$, is a parent to a random number of offspring in the $(g+1)$-st generation. However, each child node has only one parent. 

The  $k$-th realization of a tree network consists of several generations each containing a random number of nodes, 
$\D_{g,k}$, which we model by using the Galton-Watson process \cite{harris1964theory},
\begin{equation}
\label{GW.eq}
\D_{g+1,k} = \sum_{i \in \text{gen.} g} d_{g,i,k}, 
\quad g=0,1,...,G-1, \quad \D_{0,k}=1.
\end{equation}
Here $g$ indicates the generation and $k=1,2,\hdots$ indicates a particular realization of a tree network. The sum runs over all nodes in generation $g$.
The number of offspring, $d_{g,i,k}$, of a certain parent node $i$ in generation $g$ is an independent random variable generated from the probability mass function (PMF), $p_g(d)$. 
In the following sections, we will consider several examples of branching PMFs.

We consider basic properties of a single tree network realization, first. 
The total number of nodes $\N_k$ of a tree network realization $k$ is  a random variable and obtained by summing $\D_{g,k}$ over all generations,
\begin{equation}
\N_k = \sum_{g=0}^G \D_{g,k} = \sum_{g=0}^{\mathcal{G}_k} \D_{g,k},
\label{Ntotal.eq}
\end{equation}
where $\mathcal{G}_k \leq G$ is the height of the tree and is given by the actual number of generations in the current tree network realization. Thus, there might be no nodes in the outer generations for particular realizations of the Galton-Watson process, Eq. (\ref{GW.eq}). 
The number of leaf nodes in a particular generation $g$ is also a random variable given by
\begin{equation}
h_{g,k} = \sum_{i \in \text{gen.} g} \delta_{d_{g,i,k},0}.
\label{leaves_g.eq}
\end{equation} 
By construction,  branching terminates at the maximum generation $G$ and all nodes in that generation are leaf nodes. In general, however, leaf nodes can be found in any but the $0$-th generation. 
 Thus, a node $i$ in generation $0<g\leq G$ becomes a leaf node with probability $p_g(0)$, i.e. if $d_{g,i,k}=0$. 
The total number of leaf nodes, $\Nh_k$, is a random variable, obtained by summing $h_{g,k}$ over all generations, i.e.
\begin{equation}
\Nh_k = \sum_{g=1}^G h_{g,k}.
\label{mod.eqq3}
\end{equation}
In the special case of no extinction, i.e. if $p_g(0) = 0$ for $0 \le g< G$, all leaf nodes are in the peripheral generation, 
$g=\mathcal{G}_k=G$ and $\Nh_k=\D_{G,k}$. 

Next, given the branching PMF, $p_g(d)$, and the maximum allowed number of generations, $G$, some statistical properties of random trees formed by the Galton-Watson process, such as the mean number of total nodes and leaves, their variances, etc. can be determined by the standard method of probability generating functions  \cite{harris1964theory,drmota2009random,newman2010networks}. 
In the present paper, we are mainly interested in the influences of structural variability, as arising from different tree network realizations for the same branching PMF, on the firing statistics of the root node. We characterize that statistics for a single tree network realization $k$ by calculating the firing rate, $r_k(J,S,\kappa)$, and the CV, $\Cv_k(J,S,\kappa)$ according to Eqs. (\ref{mfr.eq}). Note that these quantities depend not only on the network realization with the particular coupling structure, but also on input current, i.e. $J$ and $S$, and the coupling strength $\kappa$. 

The variability of a certain quantity $Q_k$, of the $k$-th tree realization, can be assessed by calculating its ensemble average mean and standard deviation (SD):
\begin{equation}
	\mean{Q} = \lim_{K\to\infty} \frac{1}{K} \sum_{k=1}^K Q_k, \quad
	\sigma^2_Q =  \lim_{K\to\infty} \frac{1}{K} \sum_{k=1}^K Q^2_k -\mean{Q}^2.
	\label{equ:EnsembleAv}
\end{equation}
This yields the ensemble averaged firing rate $\langle r(J,S,\kappa) \rangle$ and its normalized standard deviation:
\begin{equation}
\Cv_r(J,S,\kappa) = \frac{\sigma_r(J,S,\kappa)}{\mean{r(J,S,\kappa)}}.
\label{CV_top.eq}
\end{equation}
The latter provides a measure of variability for the root nodes' firing rates resulting from tree network realizations from the same branching PMF with respective input currents and coupling strengths.

By defining sets of identical or isomorphic trees in an ensemble, averaging can be performed using the probability distribution of sets of identical trees.   
Denoting the set of parameters, which uniquely defines trees by $\{T_k\}$, the $k$-th realization of the quantity $Q$ is denoted as $Q_k := Q(\{T_k\})$. Its average can be formally written as, $\mean{Q} = \sum_{\{T\}} Q(\{T\}) \,P(\{T\})$, where $P(\{T\})$ is the PMF of non-identical trees and the summation run over all possible values of the parameters set, $\{T\}$. 
Various levels of coarse-graining methods can be used to simplify the ensemble averaging. 

In the present paper we restrict to identical nodes and interconnections except  for  the  external  input,  which  is  only  applied to  leaf  nodes. As a first way of coarse-graining, we consider trees with the same number of nodes and leaf nodes in each generation as identical.
We get $(2G-1)$-tuple $\{T_k\}=(\D_{1,k},...,\D_{G,k},h_{1,k},...,h_{G-1,k}) = (\tuple{D_k},\tuple{h_k})$. We note that trees with identical tuples $(\tuple{D_k},\tuple{h_k})$ may still possess different connectivities. The ensemble of trees is then characterized by the joint $(2G-1)$-dimensional PMF of number of the nodes and leaves in all generations, $P_{2G-1}(\D_1,...,\D_G;h_1,...,h_{G-1}) \equiv P_{2G-1}(\tuple{\D},\tuple{h})$. The tree-ensemble averages, Eqs. (\ref{equ:EnsembleAv}), are then approximated by
\begin{eqnarray}
&& \mean{Q(J,S,\kappa)} \approx \sum_{\tuple{\D},\tuple{h}} Q(J,S,\kappa,\tuple{\D},\tuple{h}) 
P_{2G-1}(\tuple{\D},\tuple{h}),\nonumber\\
&&\sigma^2_Q(J,S,\kappa) \approx \sum_{\tuple{\D},\tuple{h}} Q^2(J,S,\kappa,\tuple{\D},\tuple{h}) 
P_{2G-1}(\tuple{\D},\tuple{h}) - 
\mean{Q(J,S,\kappa)}^2,
\label{ave2.eq}
\end{eqnarray}
where the summations run over all possible values of $\D_1,...,\D_G$ and $h_1,...,h_{G-1}$. 
As a second way of coarse-graining, we consider trees with the same total number of nodes and leaf nodes as identical. As we will show in Sec.~\ref{strong_cpl.sec}, this simplification yields sufficient results in the strong-coupling limit $\kappa \rightarrow \infty$. We parameterize a particular realization of tree network by the tuples  $(\Nh_k,\N_k)$ and then carry out averaging similar to Eq. (\ref{ave2.eq}), but with 2-dimensional PMF of the total number of leaves and nodes, $P_2(\Nh,\N)$.

Throughout the paper we focus on small trees, $2<G\le 4$, with branching supported on a bounded interval. This is consistent with the topology of branched myelinated terminals of sensory neurons \cite{banks1982form,banks1997pacemaker,lesniak2014computation,marshall2016touch}. For such trees the number of configurations with distinct tuples $(\tuple{\D},\tuple{h})$ in the same ensemble, is rather small. In particular, for binary trees this enables us to list all non-identical trees in the ensemble and calculate the corresponding joint PMF. Furthermore, dynamical measures, such as the firing rate and CV, can be calculated numerically for the complete small set of trees. Thus, the structure-induced variability of these measures can be calculated according to Eq. (\ref{ave2.eq}).

\section{Results}
We use small binary trees for illustration. However, our approach is applicable to any random tree network, generated by the Galton-Watson process, as we demonstrate at the end of this section.

\subsection{Statistics of binary random trees}
In binary trees each node has at most two offspring. 
Here we consider two types of binary trees: full and non-full binary trees. The so-called full binary tree, is a tree in which every internal node has two offspring and leaves have none. In contrast, in non-full binary trees, which we term as general binary trees, the number of offspring of any internal node can be either one or two.

\subsubsection{Full binary trees}
To avoid a large number of short trees in the ensemble, we allow extinction only after $2$-nd generation. In consequence, the smallest tree possesses two generations, each with branching two. In our particular example branching after the $2$-nd generation is characterize by the PMF:
\begin{align}
\label{pg1.eq}
p_g(d) = 
\begin{cases} 
\delta_{d,2},  & 0<g\le 2 \\
p_0 \delta_{d,0} + (1-p_0) \delta_{d,2}, & 2<g<G, \\
\delta_{d,0}, & g = G.
\end{cases}
\end{align}
The resulting ensemble is parameterized by two quantities: the probability of zero branching, $p_0$, and the maximum number of generations, $G$. 
For such trees the numbers of nodes in the first two generations are fixed,
$\D_1=2$, $\D_2=4$, while the numbers of nodes in higher generations are random variables ranging from 0 to $2^g$, for $g>2$. 

The limit $p_0 \to 0$ corresponds to a tree with $\N=2^{G+1}-1$ nodes and $\Nh=2^G$ leaf nodes, located in shell $G$. In the opposite limit, $p_0\to 1$, trees are extinct after the $2$-nd generation, resulting in a  tree with 
the total number of nodes and leaf nodes $\N=7$ and $\Nh=4$, respectively. 
\begin{figure*}[ht]
	\centering
	\includegraphics[width=\textwidth]{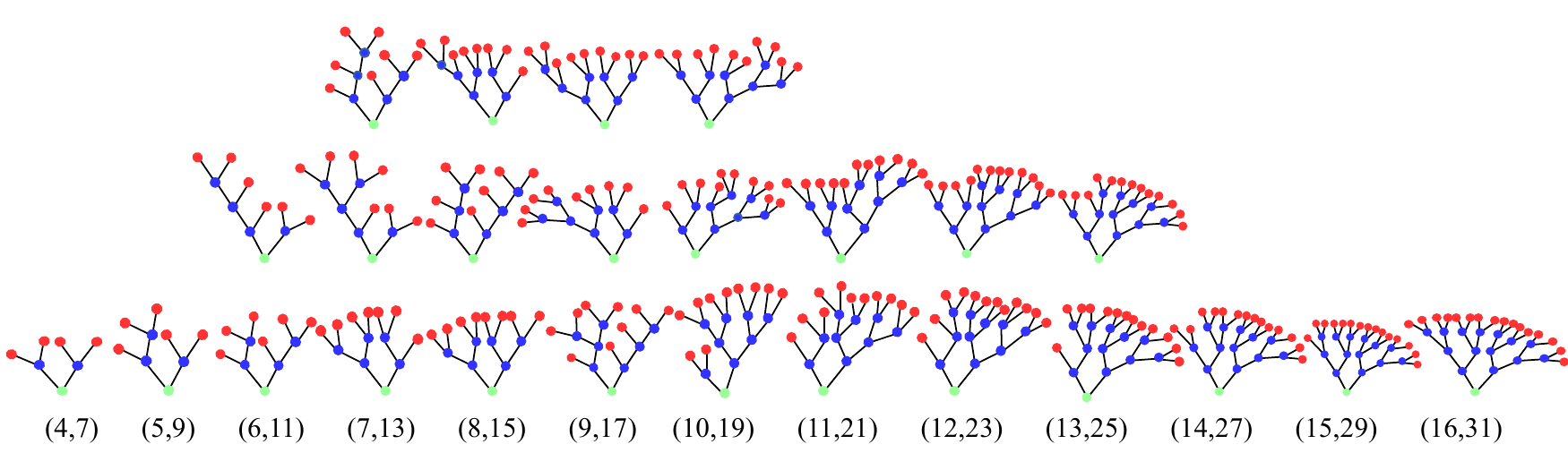}
	\caption{Tree networks for all $25$ possible configurations of $2$-tuples of $(\D_3, \D_4)$ resulting for the branching PMF given by Eq. (\ref{pg1.eq}) for $G=4$. 
		Leaves are marked red, the root node is marked  green and internal nodes are depicted as  blue circles.
		Trees are arranged in columns with the same total number of leaves and nodes, $(\Nh,\N)$, shown on the bottom.
		The ratio of the total number of nodes to leaves, $\N/\Nh$, increases from left to right, from 1.75 to 1.9375 respectively. 
	}
	\label{all_bin_trees}
\end{figure*}

\begin{figure}[h]
	\centering
	\includegraphics[width=0.7\columnwidth]{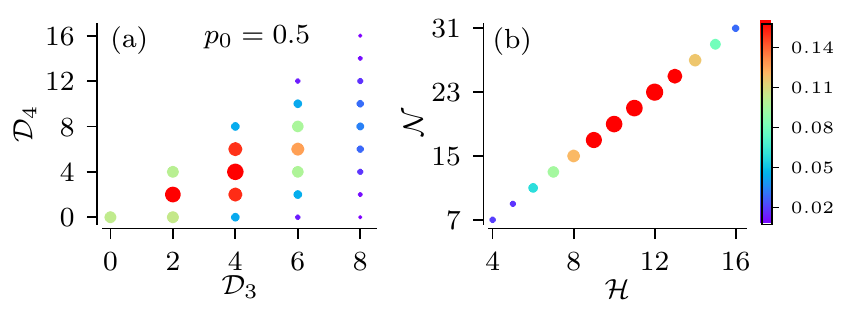}
	\caption{Node statistics of full binary trees with the branching PMF given by Eq. (\ref{pg1.eq}) with $p_0=0.5$ and $G=4$.
		(a): The joint probability mass function of numbers of nodes in $3$rd and $4$th generation,  $P_2 (\D_3,\D_4$). 
		All $25$ connectivity states are shown by filled circles. Probabilities for realizing the respective trees are indicated by circle diameter and color.
		(b): The PMF of respective combinations of total number of leaf nodes and nodes given by Eq. (\ref{jointHN}).}
	\label{jointD3D4}
\end{figure}

In the following, we consider a particular ensemble of full binary trees with at most $G=4$ generations. While the numbers of nodes in $1$-st and $2$-nd generation are fixed, the number of nodes in $3$rd and $4$th generation are random integers. The latter take even  values in the intervals $0 \le \D_3 \le 8$ and   $0 \le \D_4 \le 16$. 
Furthermore, the numbers of leaf nodes in the $2$nd and $3$rd generation are determined by the number of nodes in those shells as, $h_2=4-\D_3/2$ and $h_3=\D_3 - \D_4/2$ \cite{drmota2009random}. In consequence, trees with the same numbers of nodes in the third and fours generation can be considered as identical. This leads to 
$25$ unique $2$-tuples of $(\D_3, \D_4)$, or equivalently to $25$ distinct trees in the ensemble, shown in Fig.~\ref{all_bin_trees}. 
The joint  PMF, $P_2 (\D_3,\D_4)$, describing this ensemble, is given by (see Appendix~\ref{binary_prob.app})
\begin{eqnarray}
\label{p_d3_d4.eq}
\displaystyle
&& P_2 (\D_3,\D_4)= \displaystyle \binom{4}{n_3} \displaystyle \binom{2n_3}{n_4}
p_0^{4+n_3-n_4} (1-p_0)^{n_3+n_4}, \\
&&\D_3= 2n_3, \quad \D_4=2 n_4, \quad n_3 = 0,1,..,4, \quad n_4=0,1,...,8.\nonumber
\end{eqnarray}
This two-dimensional joint PMF is shown in Fig.~\ref{jointD3D4}a.

Of particular interest is the statistics of the total number of nodes and leaf nodes. 
As we show in  Sec.~\ref{strong_cpl.sec}, both are used to derive approximations for certain measures of spike train statistics of the root node in the strong coupling limit.  The number of distinct tuples, $(\Nh_k,\N_k)$, i.e. the number of trees with identical total numbers of leaves and nodes,
is smaller than the number of trees with identical $(\D_3,\D_4)$ tuples, as trees with different numbers of nodes in certain generations may possess the same total number nodes and leaves. This is illustrated for the example of full binary tree ensemble in Fig.~\ref{all_bin_trees}. As illustrated in the figure for $G=4$, the number of distinct tuples $(\Nh_k,\N_k)$ is $13$ and thereby smaller than the total of $25$ distinct trees in the ensemble.
Furthermore, for the considered example of full binary trees, the number of leaves is exclusively determined by the number of nodes as 
$\Nh = 4+(\N-7)/2$. Therefore, the statistics of the total number of nodes and leaves is characterized by 
the one-dimensional PMF of the total number of nodes, $P_1(\N)$ (see Appendix \ref{binary_prob.app}),
\begin{equation}
P_1(\N) =   \displaystyle \left[\sum_{n_3=0}^4 \sum_{n_4=0}^{2n_3} \delta_{n_3+n_4, m}
\displaystyle \binom{4}{n_3} \binom{2n_3}{n_4} p_0^{4+n_3-n_4} \right] 
(1-p_0)^m, \quad m=(\N-7)/2.
\label{jointHN}
\end{equation}
The PMF $P_1(\N)$ is depicted in Fig.~\ref{jointD3D4}b.

\subsubsection{General binary trees}
In general binary trees, each node has either zero, one, or two offspring. 
In our particular example, general binary trees are generated from the following branching PMF:
\begin{align}
\label{pg_3.eq}
p_g(d) = 
\begin{cases} 
\displaystyle
\frac{1}{2}\left(\delta_{d,1}+\delta_{d,2}\right),  & 0 \leq g< 1, \\
\displaystyle
p_0 \delta_{d,0} + \frac{1-p_0}{2}\left(\delta_{d,1}+\delta_{d,2}\right), & 1 \leq g < G,\\
\delta_{d,0}, & g = G.
\end{cases}
\end{align}
The root node may have either one or two offspring with equal probability. Other nodes may have either zero offspring, with probability $p_0$, or either one or two offspring, each with probability $(1-p_0)/2$. As in the case of full binary trees, an ensemble of general binary trees is parametrized by the probability of zero branching, $p_0$, and by the maximal number of generations, $G$. 
In full binary trees, the number of nodes in each generation is an even number. In contrast, for general binary trees with the branching PMF (\ref{pg_3.eq}) both odd and even number of nodes allowed. 
Compared to full binary trees, this leads to a larger number of trees with distinct tuples $(\tuple{\D},\tuple{h})$. 

\begin{figure}[h!]
	\centering
	\includegraphics[width=0.7\columnwidth]{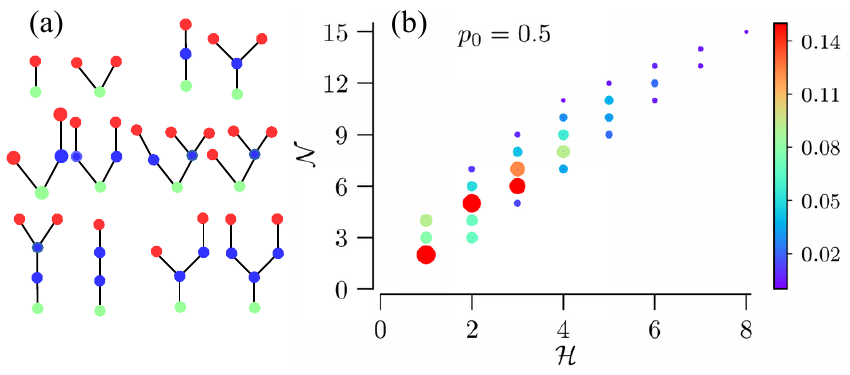}
	\caption{Realizations and statistics for general binary trees with the branching PMF (\ref{pg_3.eq}) and maximal allowed number of generation, $G=3$. 
		(a): A sample of $12$ different realizations of general binary trees.
		(b): The PMF of the total number of nodes and leaves, $P_2(\Nh, \N)$; $28$ possible trees with distinct total numbers of nodes and leaves are shown by
		filled circles, whose diameter and color represent probability values.}
	\label{trinary.fig}
\end{figure}

In order to reduce computational costs, we consider trees with the branching PMF (\ref{pg_3.eq}) and set $G=3$. 
The resulting network ensemble possesses $50$  trees with distinct sequences of numbers of nodes and leaves, $\{\D_1,\D_2,\D_3,h_1,h_2\}$. 
It is characterized by a $5$-dimensional joint PMF,
$P_5 (\D_1,\D_2,\D_3;h_1,h_2)$, which we calculated numerically for various values of the zero-branching probability $p_0$. 
Figure \ref{trinary.fig}a shows a small sample of possible tree realizations.
The limit $p_0=0$ corresponds to non-extinct binary trees, i.e. all leaf nodes are located in the $3$rd generation. The opposite limit, $p_0=1$, results in only two possible  configurations with a total number of either $2$ or $3$ nodes (one or two leaf nodes, respectively). 

In contrast to full binary trees, where the total number of nodes uniquely determines the total number of leaf nodes, general binary trees allow for several distinct configurations with the same total number of leaf nodes, but different total numbers of nodes. For the considered case of $G=3$ and $p_0 \neq 0$, the PMF given in Eq. (\ref{pg_3.eq}) leads to $28$ distinct tuples $(\Nh,\N)$, whose PMF function is illustrated in Fig.~\ref{trinary.fig}b. It shows the multiplicity of possible total numbers of nodes for the same total number of leaf nodes. 

\subsection{Strong coupling approximation}
\label{strong_cpl.sec}
In the physiologically\red{-}relevant case of strong coupling, the stochastic firing of all nodes is synchronized. 
In the synchronized state, the dynamics of individual tree network realizations, Eqs. (\ref{HHmodel.eq1}--\ref{HHmodel.eq3}), can be approximated by that of an effective single node (see Appendix~\ref{app:Averaging_Dynamics_Over_Shells}):
\begin{eqnarray}
\label{eff_node.eq}
&& C \dot{V}_k = - I_\text{ion}[V_k,m_k,h_k] + \J_{\text{eff},k}(t).
\end{eqnarray}
Here $V_k(t)$ is the effective membrane potential and  $\J_{\text{eff},k}(t)$ is the effective stochastic input current. 
The index $k$ refers to the actual tree realization. The latter encodes the tree structure.
The nodal ionic current, $I_\text{ion}[V_k,m_k,h_k]$, and the gating variables, $m_k$ and $h_k$, are given by the same equations as for the network model in Sec.~\ref{voltage_sec}.
Equations for $\J_{\text{eff},k}(t)$ for regular trees were derived in \cite{kromer2017emergent}. In Appendix~\ref{app:Averaging_Dynamics_Over_Shells}, we extend their approach to random trees and obtain,
\begin{eqnarray}
\label{Ieff.eq}
&& \J_{\text{eff},k} (t) = J_{\text{eff},k} +  \sqrt{2S_{\text{eff},k}} \, \xi(t) =  \R_{\infty,k} \, J + \sqrt{2S \Ss_{\infty,k}} \, \xi(t), \nonumber \\
&& \R_{\infty,k} = \frac{\Nh_k}{\N_k}, \quad \Ss_{\infty,k} = \frac{\Nh_k}{\N^2_k}.
\end{eqnarray}
Here $J$ and $S$ are the constant current and the noise intensity, respectively. Both specify the strengths of the noisy input current applied to the leaves in the actual tree realization; $\xi(t)$ is Gaussian white noise.

In the following, we use Eqs. (\ref{eff_node.eq}) and (\ref{Ieff.eq}) as an approximation for the dynamics of the root node, i.e. $v_k(t)\approx V_{1,k}(t)$, in the strong coupling limit. 
Thus, we replace the ensemble of trees by an ensemble of their effective root nodes. 
The respective total numbers of nodes and leaves, $(\Nh_k,\N_k)$, in the individual tree realization determine the effective current, $\R_{k} J$, and noise intensity, $S \Ss_k$, in Eq. (\ref{Ieff.eq}).

\subsubsection{Spike train statistics of tree realizations}
The strong-coupling theory allows for several important predictions.
In the strong coupling limit, the dynamics of tree nodes is determined by the total numbers of leaves and nodes, 
$\{\Nh_k, \N_k\}$, only. In consequence, it does not depend on the particular configurations with unique sequences of numbers of nodes in shells,
$\{\D_{1,k},...,\D_{G,k}\}$, on particular locations of leaves within shells,
$\{h_{1,k},...,h_{G,k}\}$, as well as on nodal connectivity.  Thus a set of distinct trees with identical total numbers of nodes and leaves, would show identical dynamics in the strong-coupling limit.

In the deterministic case, $S=0$, inputs to leaf nodes larger than the threshold current of a tree realization result in a sustained repetitive firing of the root node. 
From Eq. (\ref{Ieff.eq}), we find for the threshold current
\begin{equation}
\label{threshold_curr.eq}
J_{\infty,k} = \R^{-1}_{\infty,k} J_\text{AH},
\end{equation}
where $J_\text{AH}$ is the threshold current of a single isolated node.

\begin{figure}[h!]
	\centering
	\includegraphics[width=0.7\columnwidth]{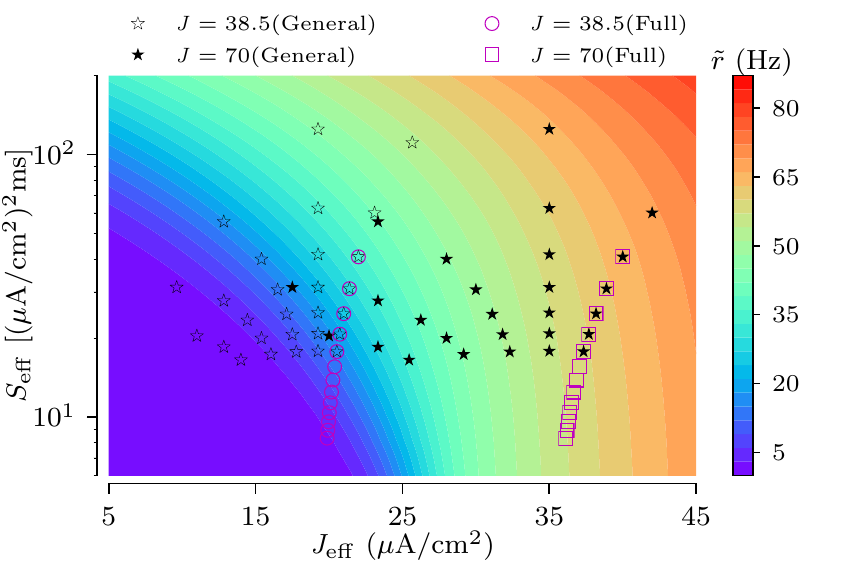}
	\caption{Heat map of the firing rate of a single HH node vs input current and noise intensity, $\widetilde{r}(\Ieff,S_\text{eff})$. 
		Symbols mark combinations of  $\Ieffk$ and $\Deffk$ (\ref{Ieff.eq}) of realizations of binary trees for the indicated ensembles. 	
		The magenta symbols (circles and squares) represent all possible full binary trees with branching PMF (\ref{pg1.eq}) and $G=4$. Black star symbols mark all possible general binary trees with branching PMF (\ref{pg_3.eq}) and $G=3$. Parameters: $S=500$~\Dunits; $J=38.5$~$\mu$A/cm$^2$ (unfilled stars and circles) and $J=70$~$\mu$A/cm$^2$ (filled stars and squares).}
	\label{heat_map.fig}
\end{figure}

In the stochastic case, the  firing statistics of the whole ensemble of trees can be predicted by evaluating the effective currents and noise intensities for all possible tuples $(\Nh_k,\N_k)$. For a single effective node, the firing rate, $\widetilde{r}(\Ieff,S_\text{eff})$, and the CV, $\widetilde{\Cv}_\tau(\Ieff,S_\text{eff})$, solely depend on these effective parameters. 
For a certain tree realization, k, in the strong-coupling limit this yields,
\begin{eqnarray}
\label{net_eq_sing.eq}
r(J,S,\kappa,\tuple{\D_k},\tuple{h_k}) \approx \widetilde{r}(\Ieffk,\Deffk),\nonumber \\
\Cv_\tau (J,S,\kappa,\tuple{\D_k},\tuple{h_k}) \approx \widetilde{\Cv}_\tau (\Ieffk,\Deffk),
\end{eqnarray}
where the tilde symbol indicates the firing rate and the CV of the effective node.  This relation is illustrated in Fig.~\ref{heat_map.fig}\red. The figure shows a heat map of  $\widetilde{r}(\Ieff,\Deff)$, where symbols mark combinations of effective parameters that can actually be realized in the presented tree network ensembles. 

\subsubsection{Spike train statistics of tree ensembles}
As follows from the previous subsection, the ensemble-averaged dynamics and its variability can be predicted from the dynamics of a single isolated node, see Eq. (\ref{eff_node.eq}--\ref{net_eq_sing.eq}), and statistics of the total numbers of nodes and leaves. This enables us to derive strong-coupling approximations for ensemble-averaged quantities such as the root node's firing rate. As the effective parameters in Eq. (\ref{eff_node.eq}) solely depend on the total number of nodes and leaf nodes, ensemble averaging in Eq. (\ref{ave2.eq}) can equivalently be performed using the two-dimensional joint PMF of total number of nodes and leaf nodes, $P_2(\Nh,\N)$. Then, the ensemble averages of the firing rates and CV can be calculated by using 
the $2$-dimensional PMF, $P_2(\N, \Nh)$, in Eqs. (\ref{ave2.eq}) as described in section \ref{stat_discr.sec} before. As we restrict on finite ranges of possible branching, it is also possible to determine bounds of corresponding quantities, such as maximal and minimal firing rates and CVs of the ISI sequence.

\subsection{Onset of repetitive spiking}
The randomness of tree ensembles may lead to qualitatively different dynamics of individual network realizations. In the present section we consider its consequence for the threshold current setting the onset of repetitive spiking of the tree's root node in the case of deterministic input currents, i.e $S=0$.

We numerically calculate the threshold current as a function of coupling strength for each of $25$ distinct full binary trees depicted in Fig.~\ref{all_bin_trees}. This results in one curve for each tree network realization, each one similar to the curve shown in Fig.~\ref{sample_tree.fig}b. Besides its dependence on the coupling strength $\kappa$, the threshold current for a single tree network realization, $J_{\text{th},k}$, is also a function of the number of nodes in the $3$-rd and $4$-th generation, i.e.  
$J_{\text{th},k}=\Ith(\kappa,\D_{3,k},\D_{4,k})$. It can be expressed in units of $J_\text{AH}$ by introducing the dimensionless scaling factor $\R^{-1}(\kappa,\D_{3,k},\D_{4,k})$, i.e. $J_{\text{th},k}=\Ith(\kappa,\D_{3,k},\D_{4,k})=\R^{-1}(\kappa,\D_{3,k},\D_{4,k}) J_\text{AH}$. At $J=J_\text{AH}$ a single isolated node enters the repetitive spiking regime by undergoing an Andronov Hopf bifurcation. 
We will refer to $\R^{-1}(\kappa,\D_{3,k},\D_{4,k})$ as the normalized threshold current in the following. For strong coupling the threshold current of a single tree network realization approaches its limiting value, $J_{\infty,k}(\Nh,\N)$, given by Eq. (\ref{threshold_curr.eq}). The latter depends only on the total number of nodes and leaf nodes, i.e.  
$J_\infty(\Nh,\N) = \R^{-1}_\infty (\Nh_k,\N_k) J_\text{AH} = (\N_k/\Nh_k) J_\text{AH}$. Here $\lim_{\kappa \to \infty} \R(\kappa,\D_{3,k},\D_{4,k}) = \R_\infty (\Nh_k,\N_k)=\Nh_k/\N_k$. 
\begin{figure}[h!]
	\centering
	\includegraphics[width=0.7\columnwidth]{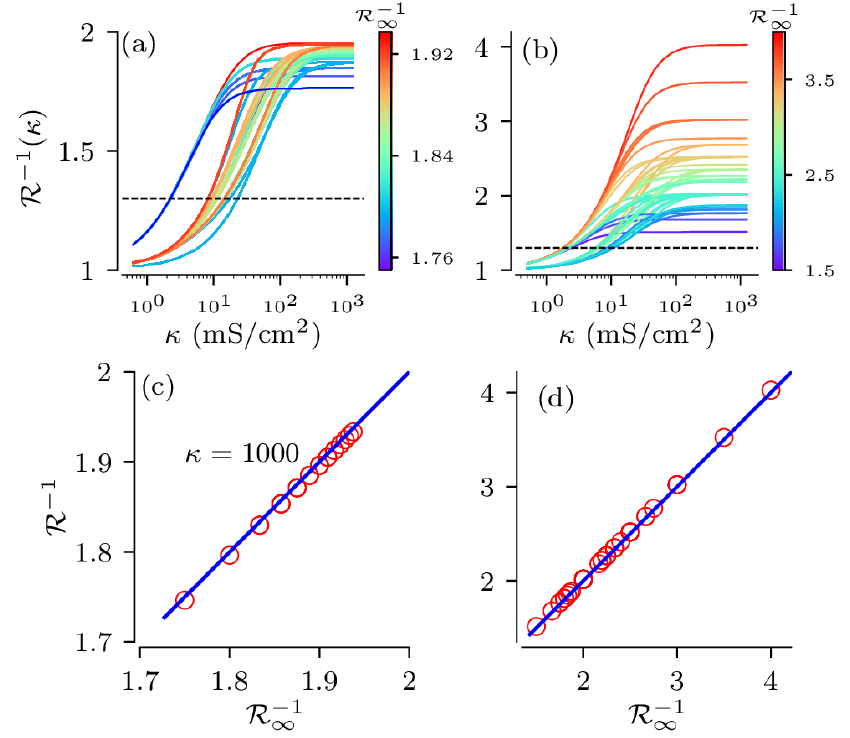}
	\caption{Normalized threshold current, $\R^{-1}(\kappa,\{\D_k\},\{h_k\}) = \Ith(\kappa, \{\D_k\}, \{h_k\})/J_\text{AH}$, as a function of coupling strength $\kappa$ for two tree network ensembles. (a) curves for all $25$ full binary trees  of Fig.~\ref{all_bin_trees}, 
		(b) same for $50$ general binary trees with all possible numbers of nodes and leaf nodes, with maximal number of generations, $G = 3$. Curves are color-coded according to the tree's normalized threshold currents in the strong coupling limit, $R^{-1}_{\infty,k}=\N_k/\Nh_k$.  Dashed horizontal lines show the value of the applied constant current, $J=38.5$~$\mu$A/cm$^2$, used for stochastic simulations in Fig.~\ref{f_cv_all_bin.fig}.  
		(c,d): Normalized threshold current for $\kappa=1000$~mS/cm$^2$ versus its theoretical strong-coupling limit,  $R^{-1}_{\infty,k}=\N_k/\Nh_k$ for full (c) and general (d) binary trees. 
	}
	\label{Ith_bin}
\end{figure}
Normalized threshold currents are shown in Fig. \ref{Ith_bin}(a,b). We find that it increases monotonically with the coupling strengths and finally saturates for strong coupling. For weak coupling, the actual tree structure matters as it determines the paths action potentials have to travel in order to excite the root node. Here  trees with various configurations, but identical numbers of leaves and nodes, fall into tight clusters around distinct values of $\R^{-1}$. With the increase of coupling the clusters blur. 
For general binary trees, Fig.~\ref{Ith_bin}(b), the range of threshold currents is significantly larger than for full binary trees, Fig.~\ref{Ith_bin}(a). 
Finally, for strong coupling curves for individual sample trees saturate. Their limiting values are well approximated by the theoretically predicted strong coupling limit $J_{\infty,k} = (\N_k/\Nh_k) J_\text{AH}$, as illustrated in Fig.~\ref{Ith_bin}(c,d). 

These results indicate that for weak and moderate coupling the onset of spike generation may be strongly affected by the particular tree structure. However, for physiologically relevant coupling strengths, $\kappa > 100$~mS/cm$^2$, threshold currents are close to their strong-coupling limits and their values mainly depend on the statistics of the total number of nodes and leaf nodes.

\subsection{Stochastic dynamics}
For stochastic input currents, i.e. $S>0$,  variability in tree structures interacts with variability caused by stochastic inputs. 
In order to study the stochastic dynamics we prepare ensembles of excitable trees, i.e. no spike generation at the root node if the input current had no stochastic component.  
To this end, we set the value of the constant input current such that it is  below the sample trees' threshold currents for strong coupling, but causes non-vanishing firing rates, $>2$~Hz, of all possible tree realizations.
For the two types of binary trees, used in the previous section, this is achieved by using a constant input current of $J=38.5~\mu$A/cm$^2$. This  corresponds to 
the normalized threshold current, $\R^{-1}=1.325$, shown by the dashed lines in Fig.~\ref{Ith_bin}a,b. For the coupling strengths, $\kappa>30$~mS/cm$^2$, all curves of the threshold current in Fig.~\ref{Ith_bin}a,b lie above the dashed lines, indicating that all trees are indeed excitable.  

Then, we simulated Eqs.(\ref{HHmodel.eq1}--\ref{HHmodel.eq3}) for all possible non-identical tree network realizations and estimated the firing rate and  CV of their root nodes, $r(\kappa,\{\D\},\{h\})$ and $\Cv_\tau (\kappa,\{\D\},\{h\})$, respectively.
\begin{figure*}[ht]
	\centering
	\includegraphics[width=1\linewidth]{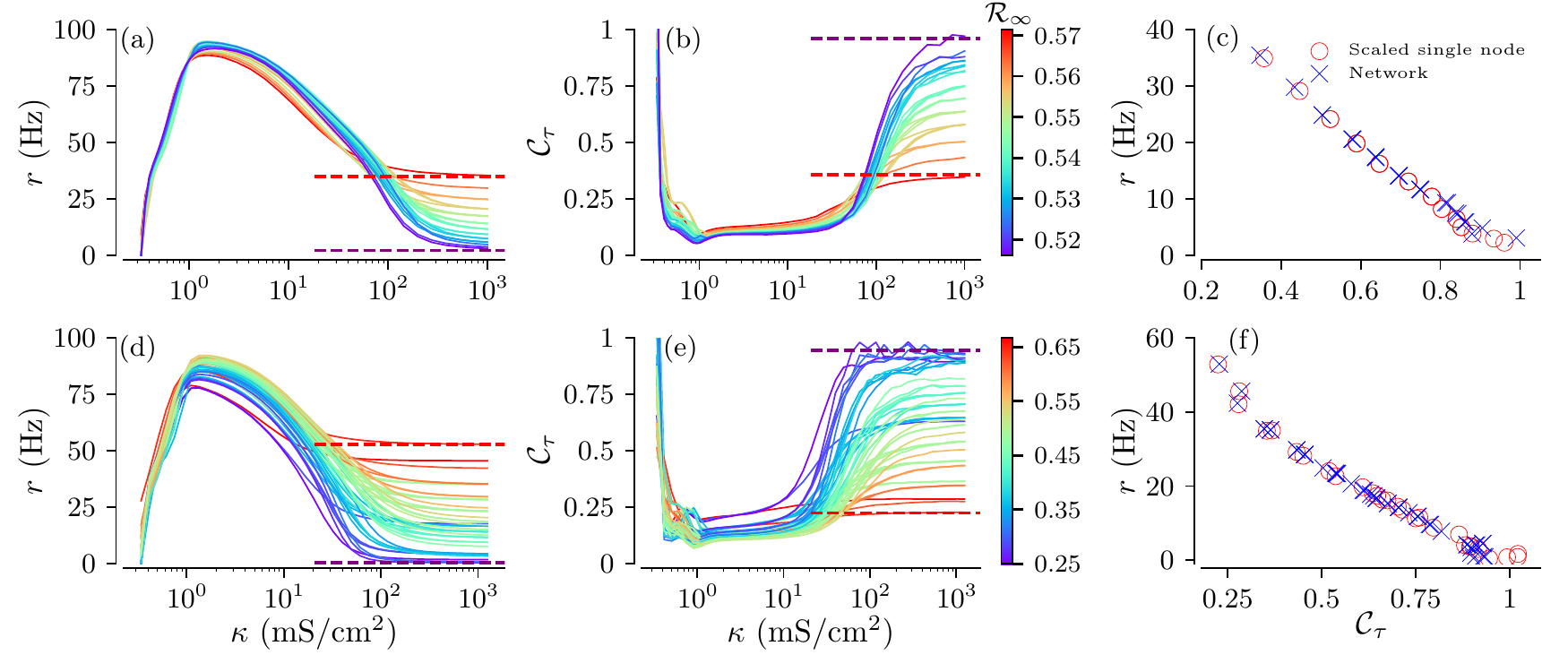}
	\caption{Spike train statistics of ensembles of excitable trees.
		The upper panels (a,b,c) show results for all $25$ full binary trees of Fig.~\ref{all_bin_trees}; the lower panels (d,e,f) refer to
		$50$ general binary trees with at most $G=3$ generations and with all possible numbers of nodes and leaf nodes.
		Trees' firing rates (a,d) and CVs (b,e) as functions of coupling strength.
		Lines are color-coded according to the scaling factor, $\R_{\infty,k}=\Nh_k/\N_k$, for the strong coupling limit; dashed horizontal lines show lower and upper limits of corresponding quantities, obtained from the strong-coupling approximation.
		(c,f): Firing rate as a function of the CV for $\kappa=1000$~mS/cm$^2$ for simulated trees (crosses) and for single root nodes (circles) with input current rescaled according to (\ref{Ieff.eq}). The parameters for numerical simulations are:
		$J=38.5$~$\mu$A/cm$^2$, $S=500$~\Dunits.
	}
	\label{f_cv_all_bin.fig}
\end{figure*}
The obtained measures of spike train statistics are shown in Fig.~\ref{f_cv_all_bin.fig}. 
For individual tree realizations, theses measures show qualitatively similar a dependences on the coupling strength. In more detail, we find that firing rates become maximal at intermediate coupling strengths for both full and general binary trees. CVs attain their minimum values at similar coupling strengths. This indicates most regular firing of the root node at intermediate coupling strengths. Note that this is in qualitative agreement with previous results on regular tree networks presented in \cite{kromer2017emergent}.  

Considering the spike train statistics of the entire tree ensemble, we find that structural variability hardly affects firing rates and CVs for weak coupling. In contrast, firing rates and CVs of individual tree realizations strongly differ for physiologically relevant strong coupling. For such coupling, the spike train statistics of  individual tree realizations are well-approximated by those of single isolated nodes with effective currents and noise intensities given by Eq. (\ref{Ieff.eq}). This is further illustrated in Fig.~\ref{f_cv_all_bin.fig}(c,f), where the  
ISIs statistics of root nodes of tree realizations is compared to those of effective isolated nodes according to the strong-coupling approximation.

Importantly, the strong-coupling approximation also allows for the prediction of lower and upper bounds of the firing rates and CVs, shown by the dashed lines in  Fig.~\ref{f_cv_all_bin.fig}a--d. The scaling parameters $\R_{\infty,k}$ and $\Ss_{\infty,k}$ of the effective current in Eq. (\ref{Ieff.eq}) define the range of the trees' firing rates and CVs at strong coupling, see Fig.~\ref{heat_map.fig}. In that sense, they provide bounds for the influence of structural variability on the spike train statistics of the root node. Since all tree realizations are excitable for strong coupling, the highest values of scaling parameters $\R_{\infty,k}$ yield the highest firing rate and lowest CV. In contrast, small values of $\R_{\infty,k}$ yield low effective currents, Eq.(\ref{Ieff.eq}), and drive the network deep into the excitable regime. This causes Poisson-like statistics of spike generation resulting in low firing rates and CVs close to one.
\begin{figure}[h!]
	\centering
	\includegraphics[width=0.7\columnwidth]{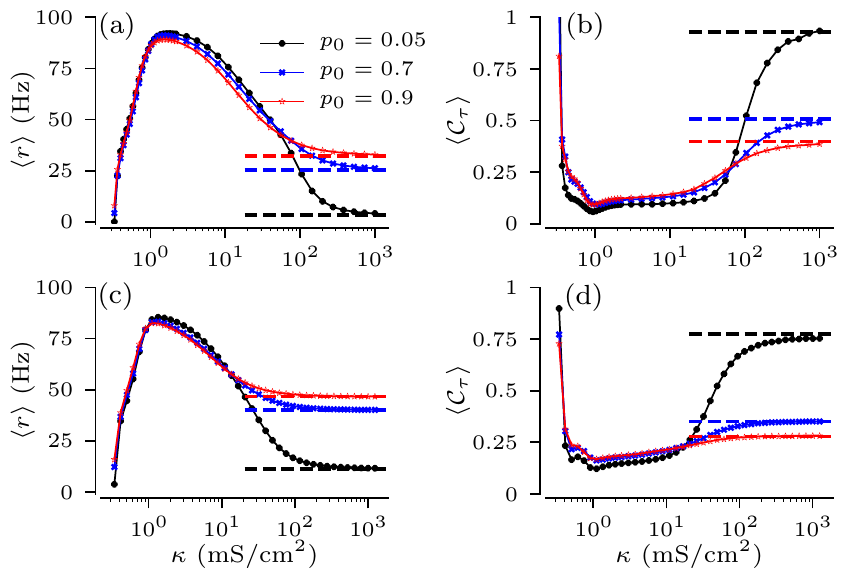}
	\caption{Ensemble-averaged ISI statistics as function of coupling strength for the indicated values of the zero-branching probability. 
		(a,b): Ensemble averaged firing rate, $\mean{r}$, and CV $\mean{\Cv_\tau}$ for full binary tress with $G=4$; 
		(c,d): $\mean{r}$ and $\mean{\Cv_\tau}$ for general binary trees with $G=3$. 
		Dashed lines show predictions of the strong-coupling theory, obtained by ensemble averaging of corresponding values for isolated single nodes with the effective current and noise intensity according to Eq. (\ref{Ieff.eq}). Parameters:
		$J=38.5$~$\mu$A/cm$^2$, $S=500$ \Dunits.}
	\label{ave_f_bin.fig}
\end{figure}

Next we consider the ensemble-averaged ISI statistics. It is obtained by averaging the firing rates and CVs  according to Eq.(\ref{ave2.eq}). 
Averaging is simplified for the full binary trees, as the joint PMF, Eq. (\ref{p_d3_d4.eq}), for $G=4$ depends only on the numbers of nodes in the $3$-rd and $4$-th generations. For the general binary trees with $G=3$ the ensemble averaging is performed using the $5$-dimensional joint PMF, $P_5 (\D_1,\D_2,\D_3;h_1,h_2)$ which we estimated numerically. 
Ensemble-averaged measures of spike train statistics are shown in  Fig.~\ref{ave_f_bin.fig} for different values of the zero-branching probability $p_0$. The ensemble-averaged firing rates and CVs follow curves that are qualitatively similar to those for individual trees. However, $p_0$ strongly affects the ensemble-averaged statistics in the strong coupling regime. Low probabilities cause on average taller sample trees with smaller fractions of leaf nodes. 
In consequence, the effective current and noise intensity in Eq. (\ref{Ieff.eq}) become smaller, which drives the network deep into the excitable regime, and results in low firing rates and large CVs. Larger values of $p_0$, refer to an increased fraction of short trees with larger fractions of leaves, which receive inputs. The latter results in higher firing rates and smaller CVs. 
\begin{figure}[h!]
	\centering
	\includegraphics[width=0.7\columnwidth]{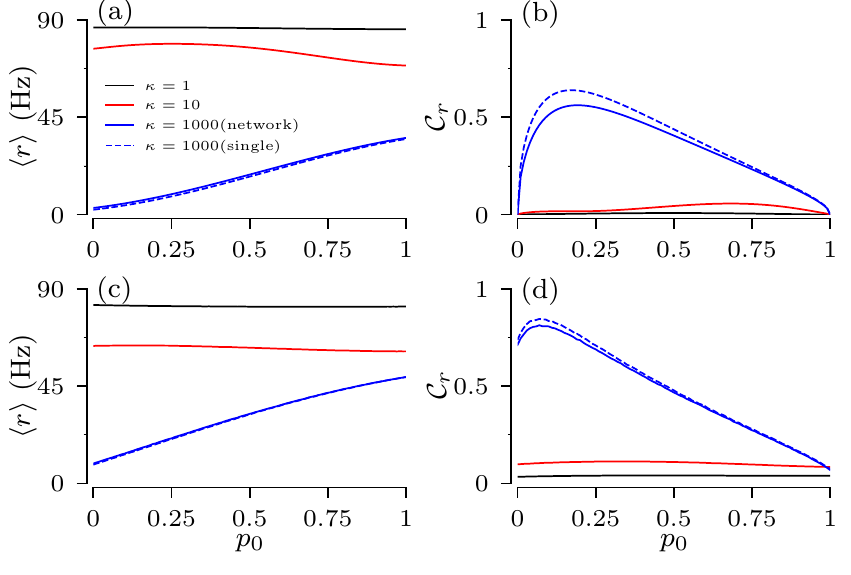}
	\caption{Variability of the firing rate due to structural variability. 
		Upper and lower panels show the ensemble averaged firing rate, $\mean{r}$, and its normalized SD,
		$\Cv_r$, (\ref{ave2.eq}, \ref{CV_top.eq}) as a function of the probability of zero branching, $p_0$, for the full binary trees and for general binary trees, respectively.
		Solid lines show results of  direct simulations of trees; dashed blue lines show prediction of the strong-coupling theory.
		Other parameters are the same as in  Fig.~\ref{ave_f_bin.fig}.
	}
	\label{ave-vs-p0.fig}
\end{figure}

In order to quantify the influence of structural variability on the firing statistics of root nodes, we consider the normalized standard deviation, $\Cv_r$, of the distribution of firing rates, see Eq. (\ref{CV_top.eq}). 
We find that firing rate distributions of general binary trees show larger variability than those for full binary trees. Besides the previously noticed fact that the structural variability is most pronounced for strong coupling, we find a  maximum of the $\Cv_r$ for a finite value of zero-branching probability. 
For full binary trees the number of possible distinct trees approaches one for $p_0 \rightarrow 0$, i.e. only the regular tree with branching two, and $p_0 \rightarrow 1$, i.e. only  the regular tree with the minimal number of generations. The pronounced maximum of  $\Cv_r$ expresses the trade-off between trees becoming more regular as $p_0$ approaches either one of those limits. In general binary trees, however, the limit $p_0 \rightarrow 0$ still results in an ensemble of 17 possible trees with distinct pairs of number of nodes and leaf nodes, as all combinations of branching one and two are  possible. Consequently, the normalized SD at $p_0=0$ remains finite.

\begin{figure}[h!]
	\centering
	\includegraphics[width=0.7\columnwidth]{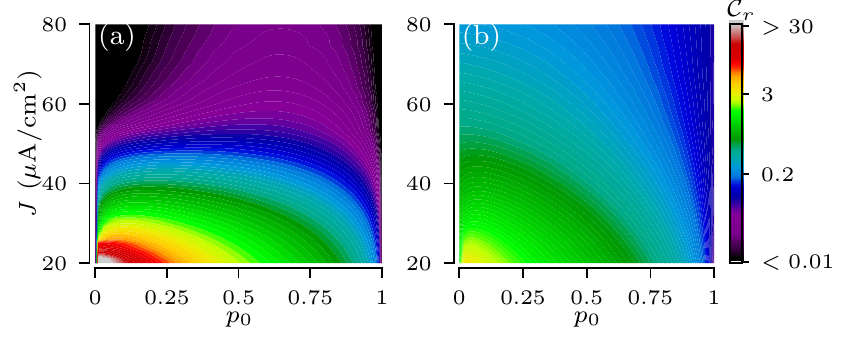}
	\caption{Normalized standard deviation of the distribution of root nodes firing rates, $\Cv_r=\Cv_r(p_0, J)$, among ensembles of full (a) and general (b) binary tries as a function of the constant input current to leaf nodes and the 
		zero-branching probability for the strong-coupling limit. Other parameters are the same as in Fig.~\ref{ave_f_bin.fig}.
	}
	\label{ave-vs-Jr.fig}
\end{figure}
The variability of the root node's firing rates indeed depends on the input current to the leaves. For strong coupling a particular coupling structure is imprinted in the scaling of input current according to Eq.(\ref{Ieff.eq}), and so the spread of effective input currents for trees in the ensemble translates into the spread of their firing rates. This  can be seen in Fig.\ref{heat_map.fig} by comparing locations of tree realizations for two values of input currents to leaves. For $J=38.5~\mu$A/cm$^2$, all tree realizations are in the excitable regime and their positions in the heat map cut across a wide range of firing rates. For $J=70~\mu$A/cm$^2$, however, most tree realizations are in oscillatory regime, cutting across a narrower range of firing rates. A decrease of the input current shifts trees to the left in 
Fig. \ref{heat_map.fig}, i.e. deeper to the excitable regime. The increase of $J$ moves trees to the right and trees become oscillatory.
As a result, the ensemble variability of firing rate is high for small input currents and low for large currents. These results are summarized  in Fig.~\ref{ave-vs-Jr.fig}, which shows the dependence of the normalized SD of firing rate, $\Cv_r$, on the input current and the zero-branching probability for the strong-coupling limit. Except for small input currents, $J < 30~\mu$A/cm$^2$, the firing rate variability of general binary trees is larger than that of full binary trees, as they cut across wider ranges of scaling factors of input current in Eq. (\ref{Ieff.eq}).

	\subsection{Non-binary trees}
	\label{non_binary.sec}
	We finally consider an example of random non-binary trees. In this example, random branching is drawn from a uniform distribution with at most $4$ offspring for each node, and  trees with at most $G=4$ generations are allowed.
	To avoid short trees we set the zero branching probability to zero, $p_g(d)=0$, for generations $g=0,1,2$, as in the example of full binary trees. Thus, the  branching PMF is given by
	\begin{align}
	\label{pg_unifrom.eq}
	p_g(d) = 
	\begin{cases} 
	\displaystyle
	\frac{1}{4}  \sum^{4}_{i=1} \delta_{d,i} ,  & 0<g\le 2, \\
	\displaystyle
	\frac{1}{5}  \sum^{4}_{i=0} \delta_{d,i}, & 2< g < 4,\\
	\delta_{d,0}, & g = 4.
	\end{cases}
	\end{align}

\begin{figure}[h]
		\centering 
		\includegraphics[width=0.7\columnwidth]{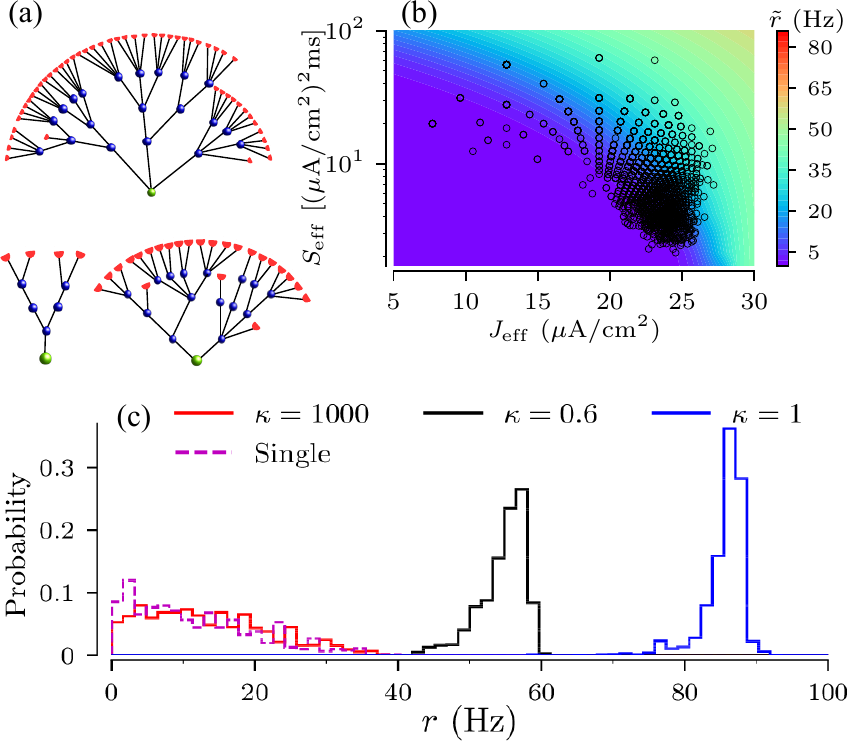}
		\caption{Firing rate statistics of non-binary random trees with uniform branching according to the PMF Eq. (\ref{pg_unifrom.eq}).
			(a) Three examples of tree networks; pairs of total  number of nodes and leaf nodes are: (49,70) for the upper panel,  (4,10) for the lower left, and (24,40) for the lower right panel, respectively. 
			(b) Heat map of the firing rate of a single HH node as a function of input current and noise intensity, $\widetilde{r}(\Ieff,S_\text{eff})$.
			Symbols mark combinations of  effective parameters, $\Ieffk$ and $\Deffk$ in Eq.(\ref{Ieff.eq}), for the ensemble of $2000$ network realizations.
			(c) Probability distributions of the firing rate of the root nodes  for the indicated values of coupling strengths.
			For strong coupling, $\kappa=1000$~mS/cm$^2$, solid and dashed lines compare  direct simulations of trees realizations and simulations of effective single nodes, respectively.
		}
		\label{uniform}
	\end{figure}

	The PMF in Eq. (\ref{pg_unifrom.eq}) yields a large number of non-identical trees, i.e. trees that differ in the their numbers of nodes and leaf nodes per generation.  
	Figure \ref{uniform}a exemplifies $3$ non-identical tree network realizations obtained from the PMF. Instead of counting distinct trees and calculating their corresponding probabilities, we analyzed tree network ensembles obtained from the PM using a brute-force approach. We generated an ensemble of $2000$ tree network realizations and then proceeded with the analyses of collective dynamics of coupled HH nodes as in previous sections.
	 
For deterministic inputs, the threshold current as a function of coupling strength possesses a similar shape as those for binary trees
(data not shown). This included the strong coupling regime, where threshold currents approached the value predicted by the strong coupling theory, $J_{\infty,k} = (\N_k/\Nh_k) J_\text{AH}$. 

	We then set the constant input current at $J=38.5$~$\mu$A/cm$^2$, for which all network realizations resulted in excitable trees for coupling $\kappa > 100$~mS/cm$^2$.
	Firing statistics of root nodes of the $2000$ tree realizations showed qualitatively similar dependences on the coupling strength as for binary trees of Fig.~\ref{f_cv_all_bin.fig} (not shown). As for binary trees, structural variability of the firing rate of non-binary trees is small for weak and intermediate coupling and large for strong coupling. This is  illustrated in Fig.~\ref{uniform}c, by the probability distribution of the firing rate of root nodes. 	The latter broadens as the coupling strength increases.
	
	As for binary trees, the firing statistics of non-binary tree network realizations can be predicted using the strong-coupling theory of Sec.\ref{strong_cpl.sec}. In order to compare the firing rate statistics obtained from simulations for the $2000$ tree network realizations with that resulting from the strong coupling approximation, we first calculate the pairs of effective current and noise intensity for each tree realization using Eq. (\ref{Ieff.eq}). The locations of those pairs in the current noise intensity space is shown in Fig. \ref{uniform}b. The figure also shows a heat map of the firing rate of a single node $\tilde{r}$ as a function of current and noise intensity. Then, using Eq. (\ref{net_eq_sing.eq}) we obtain an approximate of the firing rate for each pair of effective currents, i.e. for each tree network realization. The resulting distribution of firing rate approximates (single) is compared to the root nodes' firing rates for different coupling strengths as obtained from simulations of network activity in Fig. \ref{uniform}c.

\section{Conclusion}

We studied the influence of network structure on the root node's spike train statistics in random tree networks of excitable elements in which only the leaf nodes receive stochastic inputs.  
This setup was motivated by the morphology of certain sensory neurons which possess branched myelinated terminals with excitable nodes of Ranvier at the branch points. Myelination ends at the so-called heminodes, representing leaves of a tree, receive sensory inputs. As inputs excite heminodes, action potentials synchronously jump over myelinated branches and ultimately fire up the root node. 

Branched myelinated terminals can be represented by small random tree networks which differ in height (number of generations), numbers of nodes and heminodes, as well as nodal connectivity \cite{banks1982form,banks1997pacemaker,marshall2016touch,walsh2015mammalian}. 
Thus, the resulting spike train variability may be sensitive to the network structure.

We developed a probabilistic framework to study the collective response of stochastic excitable elements coupled on random trees whose structure is generated by Galton-Watson random branching processes. We investigated the variability of the spike train statistics resulting from variations of network structure within a  tree ensemble. We have shown that in the physiologically relevant strong coupling regime the firing statistics of the root node is determined by the number of nodes and leaf nodes, while being hardly affected by a particular nodal connectivity. Thus, trees in the ensemble can be distinguished by the total number of nodes and leaf nodes, which simplifies the calculation of the ensemble averages significantly. Furthermore, the collective response of the tree network can be predicted from a single node with an effective input, rescaled according to the number of nodes and leaf nodes. Given a joint probability distribution of the total number of nodes and leaf nodes, this allows for the calculation of the ensemble averaged firing rate and coefficient of variation as well as for setting lower and upper bounds of firing rate statistics.

Using two types of binary random trees and an example of random trees with a uniform branching we found that structural variability, resulting from different realizations of the network connectivity, strongly affects the root node's spike train statistics for strong coupling. In particular, ensembles of realizations of excitable tree networks show a wide range of firing rates and coefficients of variations, consistent with experimental findings on touch receptors \cite{wellnitz2010regularity,lesniak2014computation}.

While we considered uniform inputs to heminodes (leaf nodes),  an additional level of randomness can be introduced by non-uniform random inputs to heminodes, as in \cite{lesniak2014computation}. This would lead to additional variability across tree realizations. 
Interestingly, recent work yielded experimental evidence for structural plasticity of Merkel cell touch receptor complexes in healthy skin \cite{marshall2016touch}. The study documented that the number of heminodes of a touch receptor afferent adjusts to the inputs from Merkel cells, which varies over a time span of several days.  In our framework such an afferent remodeling corresponds to the variation of inputs, accompanied by structural changes of corresponding tree network. Strong-coupling approximation then can be used for prediction of neuronal responses during remodeling cycles.

\bigskip
\bigskip
\bigskip
\noindent
{\bf Acknowledgements}\\
The authors thank W.~Just for his valuable comments and suggestions and D.F.~Russell for discussions which inspired this work. AKN and ABN acknowledge support by the Neuroscience Program and by Quantitative Biology Institute at Ohio University. LSG thanks Ohio University for hospitality and support.

\appendix
\clearpage 

\section{Construction of Adjacency matrix}
\label{rand_adj_matrix.apdnx}
The adjacency matrix was used in numerical simulations of coupled nodes, Eqs. (\ref{HHmodel.eq1}-\ref{HHmodel.eq3}). Throughout the appendices we drop the index $k$ referring to a particular tree network realization. 
In order to construct a single tree network realization, we generate a sequence of random numbers, $\{d_g\}$, according to the branching PMF, $p_g(d)$. Then, the numbers of nodes in each generation, $\D_{0},\D_{1},\hdots,\D_{G}$ are calculated from Eq. (\ref{GW.eq}). The total number of nodes (\ref{Ntotal.eq}) 
yields the dimensions of the symmetric adjacency matrix, $\mathbf{A}$.

\begin{figure}[ht]
	\includegraphics[width=0.7\columnwidth]{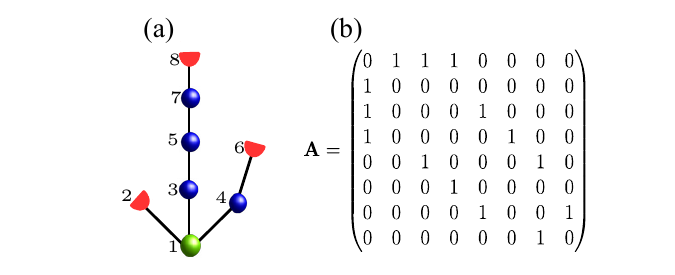}
	\caption{Example of a random tree network with $G=4$ (a) and corresponding adjacency matrix (b), constructed following Eqs. (\ref{adj1},\ref{adj2}).}
	\label{bank.fig}
\end{figure}

To construct the adjacency matrix, nodes in a tree are indexed by $j$, starting from the root node $j=1$ in the $0$-th generation and then proceeding with the nodes' offspring.  
The number of nodes until generation $g$, $\M_g$, is given by
\begin{eqnarray}
\M_g= \sum_{i=0}^{g} \D_{i},\quad g=0,\hdots,G. \nonumber
\label{Ng.eq}
\end{eqnarray}
In the first generation, the $d_{0,1}$ nodes are indexed as $j=2,...,\M_1$; nodes in the second generation are indexed in the order of their parent nodes, i.e. $j=\M_1+1, \M_1+2,..., \M_1+ d_{1,2}$ for $d_{1,2}$ offspring of node $j=2$ in generation $g=1$ and so on. This is illustrated in Fig.~\ref{bank.fig}(a). 
Thus, node indexes run from $j=1$ to  $j=\M_G=\N$. 

As the full adjacency matrix follows from symmetry, we restrict our description to the upper triangular matrix $\mathbf{A}^{\text{up}}$. Its first row contains all connections to the root node:
\begin{eqnarray}
A^{\text{up}}_{j,1}=A_{1,j}=\begin{cases}
1, \quad  j=2, \hdots, \M_1, \\
0, \quad \text{else}. 
\end{cases} 
\label{adj1}
\end{eqnarray}
Interconnections between a node $j$ in generation $g$ and its offspring in generation $g+1$ result in a sequences of ones of lengths $d_{g,j}$ in the $j$th row of $A^{\text{up}}$, where $d_{g,j}$ is the number of offspring of $j$th node, which is part of generation $g$. In more detail, 
\begin{eqnarray}
\label{eq.adj}
A^{\text{up}}_{j,i}=
\begin{cases} 
1, \quad i=(\lp+1), ~~~~~~~~~~~~~\quad j= (\M_g +1) ,\hdots \hdots \hdots . , (\M_g +d_{g,\lp+1}) \\
1,  \quad i=(\lp+2), ~~~~~~~~~~~~~\quad j= (\M_g+1+d_{g,\lp +1}) ,\hdots, (\M_g +d_{g,\lp+1}+d_{g,\lp+2}) \\
\quad \vdots ~~~~~~~~~~~~~~~~~~~~~~~~~~~~~~~~~~~~~~~~~\vdots \\
1, \quad i=(\lp+\D_{g})=\M_g,~\quad j=\big(\M_g+1+ \displaystyle \sum_{b=1}^{\D_g-1}d_{g,\lp+b}\big) ,\hdots, \big(\M_g + \sum_{b=1}^{\D_g}d_{g,\lp+b}= \M_{g+1}\big) \\

0,\quad \text{else} \nonumber
\end{cases}, 1\le g \le \mathcal{G}-1 \nonumber, \\
\label{adj2}
\end{eqnarray}

where $\lp=\M_{g-1}$. Finally, the full adjacency matrix follows from symmetry.

\section{Probability mass function of numbers of nodes and leaves for full binary trees}
\label{binary_prob.app}

For full binary trees with a maximum of $G=4$ generations and branching PMF (\ref{pg1.eq}) the joint PMF of the number of nodes and leaf nodes in 3rd and 4th generations,  $P_2(\D_3,\D_4,h_2,h_3)=P_2(\D_3,\D_4)$ is 
\begin{eqnarray}
P_2(\D_3,\D_4)= \p(\D_3) \times \p(\D_4 |\D_3),
\label{pmf1.eq}
\end{eqnarray}
where $\p(\D_3)$ is given by the binomial distribution
\begin{eqnarray}
\p(\D_3=2 n_3)={\D_2 \choose n_3} p_0^{\D_2-n_3} (1-p_0)^{n_3},
\end{eqnarray}
with integer values $n_3=0,1,..,\D_2$, specifying the number of parent nodes in generation $g=2$. Accordingly, for given $\D_3$, we find
\begin{eqnarray}
\p(\D_4=2 n_4|\D_3)={\D_3 \choose n_4} p_0^{\D_3-n_4} (1-p_0)^{n_4}.
\end{eqnarray}
Here $n_4=0,1,..,\D_3$ is the number of parent nodes in the $3$rd generation. Applying Eq. (\ref{pmf1.eq}) this yields Eq. (\ref{p_d3_d4.eq}).

For the full binary tree considered in the main text, we find $\Nh=4+(\N-7)/2$. In consequence, the joint PMF of the total number of leaf nodes and nodes is determined by the PMF of the total number of nodes, $P_1(\N)$. The latter can be obtained by summing  Eq.~(\ref{p_d3_d4.eq}) such that, $\D_3+\D_4=\N-7$, or, equivalently, $n_3+n_4=(\N-7)/2$. This yields Eq. (\ref{jointHN}).

\section{Dynamics in the strong coupling limit}
\label{app:Averaging_Dynamics_Over_Shells}
In the strong coupling limit, each of the network realizations approaches a synchronized state. Its dynamics can be treated as that of a single node given by Eq. (\ref{eff_node.eq}). In the following, we derive the effective parameters, see Eq. (\ref{Ieff.eq}), which account for the influence of network structure on the dynamics in the strong coupling limit. In order to simplify the notations, we skip the index referring to the considered network realization.

For the derivation, we follow the approach presented in \cite{kromer2017emergent}. We first consider the coupling term in Eq. (\ref{HHmodel.eq1}). Instead of using the adjacency matrix, (see Appendix \ref{rand_adj_matrix.apdnx}), we can rewrite the coupling term as a sum over interconnections between adjacent nodes:
\begin{eqnarray}
\label{cable1}
C \dot{V_j}&=& - I_{\text{ion},j}+ \J_j(t) + \kappa \sum_{i \in \text{gen. }g_j-1} A_{j,i}\bracket{V_i-V_j} + \kappa \sum_{m \in \text{gen. }g_j+1} A_{j,m}\bracket{V_m-V_j}\\  
&=&- I_{\text{ion},j}+ \J_j(t) + \kappa (1-\delta_{g_j,0})\bracket{V_{m_j}
	-V_j} + \kappa \sum_{{o_j} \in \,{\text{offspring of} \,j}}
\bracket{V_{o_j}-V_j}.
\end{eqnarray} 
Here $g_j$ denotes the generation of node $j$.  In the first line, the sums run over all nodes in adjacent generations.  In the second line, only nodes that are connected to node $j$ are considered, i.e. its only parent node, $m_j$, in generation $g_j-1$ and all offspring in generation $g_j+1$. The term with the Kronecker delta accounts for the fact that the root node has no parent. Note that the relation between nodes, their offspring, and their parent nodes implies that $k=o_j \iff j=m_k$.

\begin{figure}[h!]
	\centering \includegraphics[width=0.5\linewidth]{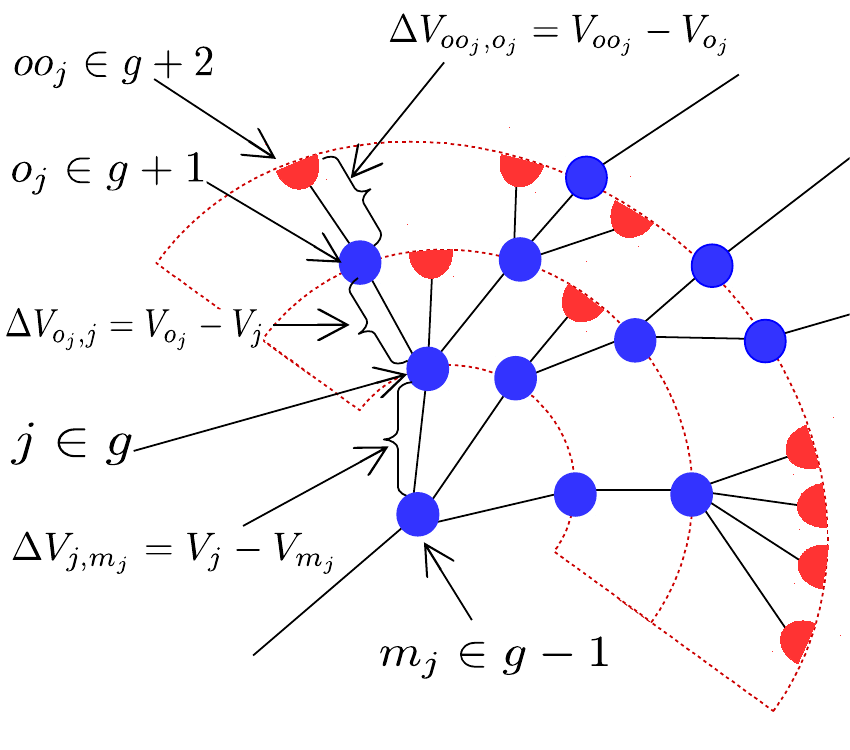}
	\caption{Illustration of notations for the voltage differences, $\Delta
		V_{j,m_j}$, $\Delta V_{o_j,j}$, and $\Delta
		V_{oo_j,o_j}$ for a tree fragment.  Therein $o_j \in (g+1)$ labels  the node under consideration; the node with the index $j \in g$ is the parent of $o_j$. The node with the index $o_j \in (g+1)$ is the offspring of $j$ and is the parent of the node $oo_j \in
		(g+2)$. Red semicircles show leaves at which branching is terminated.}
	\label{ff1}
\end{figure}
As only the differences in membrane potentials between nodes and their offspring enter Eq. (\ref{cable1}), we introduce these differences as new variables
\begin{eqnarray}
\Delta V_{o_j,j}:= V_{o_j} - V_j.
\end{eqnarray}
Voltage difference and the corresponding notation is illustrated in Fig.~\ref{ff1}. 
Applying this to Eq. (\ref{cable1}), we obtain 
\begin{eqnarray}
C \dot{V_j}= - I_{\text{ion},j} + \J_j(t) - \kappa (1-\delta_{g_j,0}) \Delta V_{j, m_j} + \kappa \sum_{o_j \in \text{offspirng of }\,j} \Delta V_{o_j, j}.  
\end{eqnarray}
From this, we can derive the dynamics of the voltage differences by subtraction of the two equations for $V_{o_j}$ and $V_j$:  
\begin{eqnarray}
C \frac{{\rm d}}{{\rm d}t}\Delta{V}_{{o_j},j} &=& - I_{\text{ion},o_j}+ I_{\text{ion},j} + \J_{o_j}(t) - \J_j(t) +
\kappa (1-\delta_{g_j,0}) \Delta V_{j,\, m_j}\notag \\
&&- \kappa \bracket{ \Delta
	V_{{o_j},j} + \sum_{o_j^\prime \in \text{offspring of }\,j}
	\Delta V_{o_j^\prime,j}} + \kappa \sum_{oo_j^{\prime} \in 
	\text{offspring of }\,o_j} \Delta V_{oo_j^{\prime},o_j}
\end{eqnarray}

\subsection{Generation-averaged dynamics}

In the strong coupling limit, the nodal dynamics is synchronized and nodes within the same generation become statistically indistinguishable.
To make use of this fact, we follow the approach presented by Kouvaris et al. \cite{ko2014} and extend it to stochastic excitable elements on random tree networks.
To this end, we consider the dynamics of the generation-averaged membrane potentials 
\begin{eqnarray}
\label{equ:shellAvMembranePotential}
\av{V}_g := \frac{1}{\D_g} \sum \limits_{j \,\in \, \text{gen.} g}
V_j,~~~~ g=0,1,2,\ldots, \mathcal{G}.
\end{eqnarray}
Here the average is taken in
each generation $g<\mathcal{G}$ of the $k$th random tree network realization (the realization index $k$ is dropped), i.e. only generation that actually include nodes are considered. 

Applying Eq. (\ref{equ:shellAvMembranePotential}), to Eq. (\ref{cable1}), we obtain the generation-averaged membrane potential dynamics
\begin{equation}
\label{cable3}
C \frac{{\rm d}}{{\rm d}t} \av{V}_{g}= - \av{I_\text{ion}}_{g} + \av{\J(t)}_{g} - \kappa
\av{V}_{g}+ \kappa \frac{1}{\D_{g}} \sum_{j \in \text{gen.} g} V_{m_j}
+ \kappa \frac{1}{\D_{g}} \sum_{j \in \text{gen.} g} \,\,\sum_{o_j  \in
	\text{offspring of }\, j} \bracket{V_{o_j}-V_j}.
\end{equation}
Note that averaging over the zeroth generation yields $\av{V}_0=V_1$.
As the first sum on the right-hand side of Eq. (\ref{cable3}) runs over the $\D_{g}$ offspring which share $\D_{g-1}$ parent nodes, we find
\begin{equation}
\label{eq:D8}
\frac{1}{\D_{g}} \sum_{j \in \text{gen. }g} V_{m_j}=\frac{\D_{g-1}}{\D_{g}}
\frac{1}{\D_{g-1}}\sum_{i \in \text{gen. }g-1} d_{g-1,i} \,V_i=
\frac{\D_{g-1}}{\D_{g}} \av{d \, V}_{g-1}.
\end{equation}
Accordingly, the double sum in (\ref{cable3}) can be simplified by noting that the $j$th node in generation $g$ has $d_{g,j}$ offspring  in generation $g+1$, 
\begin{equation}
\label{eq:D9}
\frac{1}{\D_{g}} \, \sum_{j \in \text{gen. }g} \,\sum_{o_j \in \text{offspring of
	}\,j} \,V_j\, = \,\frac{1}{\D_{g}} \sum_{j \in \text{gen. } g} d_{g,j} V_j =
\av{d \, V}_{g}.
\end{equation}
The other double sum can be reduced to a single sum over the nodes in the $(g+1)$th  generation
\begin{equation}
\label{eq:D10}
\frac{1}{\D_{g}} \sum_{j \in g} \,\,\sum_{o_j \in \text{gen. } g+1} V_{o_j}\,=\,
\frac{1}{\D_{g}} \sum_{o_j \in \text{gen. }g+1} V_{o_j}. \nonumber
\end{equation}	
This can be further simplified by considering the generation average membrane potential, 
\begin{equation}
\label{eq:D11}
\frac{1}{\D_{g}} \sum_{o_j\in g+1}
V_{o_j}\,=\,\frac{\D_{g+1}}{\D_{g}}
\,\frac{1}{\D_{g+1}}\,\sum_{o_j \in g+1} V_{o_j}
=\,\frac{\D_{g+1}}{\D_{g}} \,\av{V}_{g+1}.
\end{equation}
From the recurrent relation of the Galton-Watson process (\ref{GW.eq}) it follows that
the ratios of numbers of nodes in adjacent generations can be replaced by the mean branching within a generation as
$\D_{g+1}/\D_{g}=\av{d}_{g}$. 
Then using Eqs.(\ref{eq:D8} -- \ref{eq:D11}) in Eq. (\ref{cable3}) yields 
\begin{equation}
C \frac{{\rm d}}{{\rm d}t} \av{V}_{g}= - \av{I_\text{ion}}_{g} +  \av{\J(t)}_{g} - \kappa
\bracket{\av{V}_{g}\,-\,\frac{1}{\av{d}_{g-1}} \av{d\,V}_{g-1}}
\,+\,\kappa \bracket{\av{d}_g 	\,\av{V}_{g+1}\,-\,\av{d\,V}_{g}}.
\end{equation}

In the limit of strong coupling, we can assume that {$V_i \approx \av{V}_{g,i}$ for $i$ in generation $g_i$. Then, we can decouple the branching from the generation average, i.e. $\av{d\,V}_{g} \approx \av{d}_{g}\,\av{V}_{g}$.  Performing similar simplification for the root node, $g=0$ and the nodes in the last generation $g=G$, we end up with the following system for the generation-averaged membrane potentials
\begin{eqnarray}
\label{eq:shellAvDynamics}
C \frac{{\rm d}}{{\rm d}t} \av{V}_{0}\,&=&\, - \av{I_\text{ion}}_{0} + \av{\J(t)}_{0} \,+ \,\kappa \
d_{0,1} \bracket{ \,\av{V}_{1}\,-\, \av{V}_{0}}\, ,  \notag\\
C \frac{{\rm d}}{{\rm d}t} \av{V}_{g}&=& - \av{I_\text{ion}}_{g} + \av{\J(t)}_{g} - \kappa
\bracket{\av{V}_g\,-\, \av{V}_{g-1}} \,+\,\kappa \av{d}_g
\,\bracket{\av{V}_{g+1}\,-\,\av{V}_{g}}\,,\,\,\,g=1,\ldots,\mathcal{G}-1,\notag\\
C\frac{{\rm d}}{{\rm d}t}\av{V}_{\mathcal{G}}\,&=&\,- \av{I_\text{ion}}_{\mathcal{G}} + \av{\J(t)}_{\mathcal{G}} + \kappa \bracket{
	\av{V}_{\mathcal{G}-1}\,-\,\av{V}_{\mathcal{G}} }. 
\end{eqnarray}

\subsection{Dynamics of generation-averaged membrane potential differences}

Next we introduce the differences between generation-averaged membrane potentials from adjacent generations, $\Delta \av{V}_{g} :=
\av{V}_{g+1}-\av{V}_{g}$, $g=0,1,\ldots,\mathcal{G}-1$. Subtraction of the corresponding equations in (\ref{eq:shellAvDynamics}) yields
\begin{eqnarray}
\label{eq_main}
C \frac{d}{dt}\av{\Delta V}_{g}&=&-\Delta \av{I_\text{ion}}_{g}  +  \Delta \av{\J(t)}_{g} +\notag \nonumber\\
&+&\begin{cases}  - \kappa \,(\av{d}_{0}+1) \Delta \av{V}_{0} +  \kappa \av{d}_{1} \Delta \av{V}_{2},  & g=0,\\
& \\
\kappa \Delta \av{V}_{g-1}-\kappa \bracket{\av{d}_{g} +1} \Delta \av{V}_{g} + \kappa \av{d}_{g+1} \Delta \av{V}_{g+1}, & 1\leq g<\mathcal{G}-1, \\
& \\
\kappa \ \Delta \av{V}_{\mathcal{G}-2}  - \av{d}_{\mathcal{G}-1}\Delta \av{V}_{\mathcal{G}-1},  & g=\mathcal{G}-1, \\
\end{cases}\notag \\
\end{eqnarray}
Here we introduced the differences between generation-averaged ionic currents $\Delta \av{I_\text{ion}}_{g} := \av{I_\text{ion}}_{g+1}-\av{I_\text{ion}}_{g}$ and input currents $\Delta \av{\J(t)}_{g}=\av{\J(t)}_{g+1}-\av{\J(t)}_{g}$ and set 
$d_{0,1}=\av{d}_{0}$. The differences of generation-averaged input currents $\Delta \av{\J(t)}_{g}$ can be separated in a deterministic part $\Delta \av{J}_{g} := \av{J}_{g+1}-\av{J}_{g}$ and a stochastic one $\Delta \xi_{g}(t) = \av{\xi(t)}_{g+1}-\av{\xi(t)}_{g}$. 

Next, we consider the difference of the generation-averaged ionic currents $\Delta \av{I_\text{ion}}_{g}$. Both, the difference between membrane potentials of individual nodes and corresponding generation averages, and the difference between generation-averaged membrane potentials of adjacent generations $\Delta \av{V}_{g}$ become small in the case of strong coupling. We therefore approximate the differences between generation-averaged ionic currents to the first order in $\Delta \av{V}_{g}$ around a vanishing mean difference. We assume that it can be expanded in a Taylor expansion
around $\Delta \av{V}_{g}=0$. 
This yields $\av{I_\text{ion}}_{g+1} - \av{I_\text{ion}}_{g} 
\approx a_g + b_g \Delta \av{V}_g + h.o.$. As we restrict on networks of identical nodes, except for leaf nodes, we assume that the coefficient $a_g$ vanishes and that the coefficients $b_g$  are small compared to the coupling strength $\kappa$, i.e. $b_g \ll \kappa$. Using these assumptions, Eq. (\ref{eq_main}) can be linearized and we find
\begin{eqnarray}
\label{eq_main1}
C\frac{d}{dt}\av{\Delta V}_{g,k}&\approx & \Delta \av{\J(t)}_{g}+ \nonumber\\ &+&\begin{cases} - \kappa
\bracket{ \av{d}_{0}+1} \Delta \av{V}_0 + \kappa \av{d}_{1}
\Delta \av{V}_{1}, & g=0,\\ & \\ \kappa \Delta
	\av{V}_{g-1}-\kappa \bracket{\av{d}_{g} +1}
\Delta \av{V}_g + \kappa \av{d}_{g+1} \Delta \av{V}_{g+1}, &
1 \leq g<\mathcal{G}-1, \\ & \\ \kappa \Delta \av{V}_{\mathcal{G}-2} -
\kappa \bracket{\av{d}_{\mathcal{G}-1}+1}\Delta
	\av{V}_{\mathcal{G}-1}, & g=\mathcal{G}-1, \\
\end{cases}\notag \\
\end{eqnarray}
In consequence, the dynamics of the differences of the generation-averaged membrane potentials can be
approximated by a multidimensional Ornstein-Uhlenbeck process,
\begin{eqnarray}
\label{cable6}
C\frac{d}{dt}\boldsymbol{\Delta} \av{ \textbf{V}} \approx
\mathbf{B} \boldsymbol{\Delta}\av{\textbf{V}} +
\boldsymbol{\Delta} \av{\textbf{J}} + \boldsymbol{\Delta}\av{\boldsymbol{
	\xi}}(t).
\end{eqnarray}
Here we introduced the $G$-dimensional vectors,
\begin{eqnarray}
&&\boldsymbol{\Delta} \av{\textbf{V}}=(\Delta \av{V}_{0},..., 
\Delta \av{V}_{\mathcal{G}-1})^T, \nonumber \\
&& \boldsymbol{\Delta}
\av{\textbf{J}}=(\Delta \av{J_{0}},\Delta \av{J_{1}},..., \Delta
\av{J_{\mathcal{G}-1}})^T, \nonumber \\
&&\boldsymbol{\Delta}
\av{\boldsymbol{\xi}(t)}=\left(\Delta \av{\xi(t)}_{0},\Delta
\av{\xi(t)}_{1},...,\Delta \av{\xi(t)}_{\mathcal{G}-1} \right)^T,\nonumber
\end{eqnarray}
and the $\mathcal{G} \times \mathcal{G}$ tridiagonal matrix,
\begin{eqnarray}
\label{eq:matrixBG} 
\mathbf{B}= \begin{pmatrix}
-\kappa \bracket{\av{d}_{0}+1} & \kappa \av{d}_{1} & 0 & ... & 0 \\
\kappa  & -\kappa \bracket{\av{d}_{1}+1} & \kappa  \av{d}_{2} & ... & ... \\
0 & \kappa  & -\kappa  \bracket{\av{d}_{2}+1} & ... & 0 \\
... & ... & ... & ... & \kappa \av{d}_{\mathcal{G}-1} \\
0 & .. & 0 & \kappa  & -\kappa \bracket{\av{d}_{\mathcal{G}-1}+1} \\
\end{pmatrix}.
\end{eqnarray} 
In accordance to our notation for differences of generation-averaged quantities, we introduced the differences of generation-averaged constant and noisy current components, $\Delta \av{J}_{g}=\av{J}_{g+1}-\av{J}_{g}$ and $\Delta \av{\xi(t)}_{g}=\av{\xi(t)}_{g+1}-\av{\xi(t)}_{g}$.

In the strong coupling limit, temporal deviations of
$\boldsymbol{\Delta} \langle{ \textbf{V} }\rangle$ from its stationary value
decay extremely fast. Hence, we can use an adiabatic approximation
\cite{vankampen1985},  to approximate
$\boldsymbol{\Delta} \av{\textbf{V}}$ by its stationary value plus a
white Gaussian noise. Both, the stationary voltage difference and the
intensity of the Gaussian white noise in the strong coupling limit can
be obtained by setting the left-hand side of Eq. (\ref{cable6}) to
zero. This yields
\begin{equation}
\label{eq:DensityDifferences}
\boldsymbol{\Delta}\av{ \textbf{V}} \approx - \mathbf{B}^{-1}
\left( \boldsymbol{\Delta} \av{\textbf{J}} +
\boldsymbol{\Delta} \av{\boldsymbol{\xi}(t)} \right),
\end{equation}
where $\mathbf{B}^{-1}$ is the inverse of the matrix $\mathbf{B}$.

\subsection{Single node description for strongly-coupled random tree networks}

In order to obtain an approximation for the dynamics of the root
node, only the first component, $\Delta \av{V}_{0}$, of Eq. (\ref{eq:DensityDifferences}) is need. Using the latter in Eq. (\ref{eq:shellAvDynamics}) yields
\begin{eqnarray}
\label{eq:singleNodeStep1}
C\dot{V}_{1} =-
I_{\text{ion},1} + \kappa \ d_{0} \bracket{\mathbf{B}^{-1} \bracket{\boldsymbol{\Delta}
		\av{ \textbf{I} } + \boldsymbol{\Delta} \av{ \boldsymbol{\xi}(t)} } }_1.
\end{eqnarray}
Hereafter the index "$1$" denotes the first component
of a $G$-dimensional vector. From Eq.(\ref{eq:singleNodeStep1}), we find the effective current $J_{\text{eff}}$ and noise intensity 
$S_{\text{eff}}$ for the current realization of the tree network as
\begin{eqnarray}
\label{eq:GeneralEffCurrentNoiseInt}
J_{\text{eff}}= \kappa  d_{0} \left( \mathbf{B}^{-1} \boldsymbol{\Delta}
\av{\textbf{J}} \right)_1, \quad \sqrt{2 S_{\text{eff}}}\, \xi(t)=\kappa \  d_{0} \left(
\mathbf{B}^{-1} \boldsymbol{\Delta}\av{\boldsymbol{\xi}(t)} \right)_1.
\end{eqnarray}
The later relation is obtained by noting that the sum of Gaussian white noises yields a Gaussian white noise with modified intensity.

For a given tree realization the inverse of $\mathbf{B}$ can be calculated explicitly using the formula for the inverse matrix
\begin{eqnarray}
\mathbf{B}^{-1}=\frac{1}{| \mathbf{B}|} \ \text{adj}(\mathbf{B}).
\end{eqnarray}
Here $| \mathbf{B}|$ and $\text{adj}(\mathbf{B})$ refer to the determinant and adjugate of the matrix $ \mathbf{B}$, respectively. 
In the following, we present explicit formulas for the cases used in the main text, $G=3,4$.

In case of $G=3$, $\mathbf{B}_k$ is a $3 \times 3$ matrix. Its determinant reads
\begin{equation}
|\mathbf{B}|\,=- \kappa^3 \bracket{\av{d}_{0} \av{d}_{1} \av{d}_{2}+\av{d}_{0}\av{d}_{1}+\av{d}_{0} +1} = - \kappa^3 \N. 
\end{equation}
Its adjugate matrix reads
\begin{eqnarray}
\text{adj} \ \mathbf{B}= \kappa^2 \begin{pmatrix}
\av{d}_{1}\av{d}_{2} + \av{d}_{1} + 1 &  \av{d}_{2}+1 & 1 \\
\av{d}_{1}\av{d}_{2} + \av{d}_{1} & \av{d}_{0}\av{d}_{2} + \av{d}_{0} + \av{d}_{2} +1 & \av{d}_{0} + 1 \\
\av{d}_{1}\av{d}_{2} & \av{d}_{0}\av{d}_{2}+\av{d}_{2}  & \av{d}_{0}\av{d}_{1}+\av{d}_{0}+1 \\
\end{pmatrix}^{\text{T}}.
\end{eqnarray}
Using this, we can evaluate the effective parameters in Eq. (\ref{eq:GeneralEffCurrentNoiseInt}) and find
\begin{equation}
\label{eq:EffectiveParameters}
J_{\text{eff}}=\mathcal{R}_{\infty} J=\,\frac{\mathcal{H}}{\mathcal{N}}\,J, \ \ \ S_{\text{eff}}=\mathcal{S}_{\infty}S= \,\frac{\mathcal{H}}{\mathcal{N}^2}\,S.
\end{equation}
Similarly, in the case of $G=4$,  $\mathbf{B}$ is a $(4 \times 4)$-matrix with determinant $|\mathbf{B}|= - \kappa^4 \N$. Evaluation of the adjugate matrix, also yields Eq.(\ref{eq:EffectiveParameters}). 

We stress that derivations in this appendix are done for the particular tree realization. Thus, assigning the  index $k$ for tree realizations in (\ref{eq:EffectiveParameters}) to the total number of leaves and nodes gives  the scaling relations Eq.(\ref{Ieff.eq},\ref{threshold_curr.eq}) of the main text.

\newpage

%

\end{document}